\documentclass[fleqn,usenatbib]{mnras}
\usepackage{newtxtext,newtxmath}
\usepackage[T1]{fontenc}
\usepackage{ae,aecompl}
\usepackage{graphicx}	
\usepackage{amsmath}	
\usepackage{amssymb}	
\usepackage[flushleft]{threeparttable}
\newcommand{\swift}{{\it Swift}}
\newcommand{\lco}{{LCO}}
\newcommand{\LCO}{{LCO}}
\newcommand{\fair}{Fairall~9}
\newcommand{\msun}{$\mathrm{M}_{\odot}$}

\newcommand{\orcid}[1]{\textsuperscript{\href{http://orcid.org/#1}{
\hskip2pt\includegraphics[width=8pt]{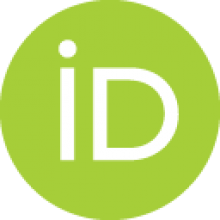}}}}

\title[Continuum echo mapping of \fair]{Intensive disc-reverberation mapping of \fair: 1st year of \textit{Swift} \& \LCO\ monitoring} 
\author[Hern\'andez Santisteban et al.]{J.~V. Hern\'andez Santisteban\orcid{0000-0002-6733-5556},$^{1}$\thanks{E-mail: JVHS (jvhs1@st-andrews.ac.uk)}
R. Edelson\orcid{0000-0001-8598-1482},$^{2}$
K. Horne\orcid{0000-0003-1728-0304},$^{1}$
J.~M. Gelbord\orcid{0000-0001-9092-8619},$^{3,4}$
\newauthor
A.~J. Barth\orcid{0000-0002-3026-0562},$^{5}$
E.~M. Cackett\orcid{0000-0002-8294-9281},$^{6}$
M.~R. Goad,$^{7}$
H. Netzer\orcid{0000-0002-6766-0260},$^{8}$
D. Starkey,$^{1}$
\newauthor
P. Uttley\orcid{0000-0001-9355-961X},$^{9}$
W.~N. Brandt\orcid{0000-0002-0167-2453},$^{10,11,12}$
K. Korista\orcid{0000-0003-0944-1008},$^{13}$
A.~M. Lohfink,$^{14}$
C.~A. Onken\orcid{0000-0003-0017-349X},$^{15,16}$
\newauthor
K.~L. Page\orcid{0000-0001-5624-2613},$^{7}$
M. Siegel,$^{10}$
M. Vestergaard\orcid{0000-0001-9191-9837},$^{17}$
S. Bisogni\orcid{0000-0003-3746-4565},$^{18}$
A.~A. Breeveld\orcid{0000-0002-0001-7270},$^{19}$
\newauthor
S.~B. Cenko,$^{20,21}$
E. Dalla Bont\`a\orcid{0000-0001-9931-8681},$^{22,23}$
P.~A. Evans\orcid{0000-0002-8465-3353},$^{7}$
G. Ferland\orcid{0000-0003-4503-6333},$^{24}$
\newauthor
D.~H. Gonzalez-Buitrago\orcid{0000-0002-9280-1184},$^{25}$
D. Grupe\orcid{0000-0002-9961-3661},$^{26}$
M.~D. Joner\orcid{0000-0003-0634-8449},$^{27}$
G. Kriss\orcid{0000-0002-2180-8266},$^{28}$
\newauthor
S.~J. LaPorte,$^{29}$
S. Mathur\orcid{0000-0002-0129-0316},$^{30,31}$
F. Marshall,$^{32}$
M. Mehdipour\orcid{0000-0002-4992-4664},$^{33}$
D. Mudd\orcid{0000-0003-2371-4121},$^{5}$
\newauthor
B.~M. Peterson\orcid{0000-0001-6481-5397},$^{30,31,28}$
T. Schmidt\orcid{0000-0002-2772-8160},$^{5}$
S. Vaughan\orcid{0000-0003-4808-092X},$^{7}$
and S. Valenti\orcid{0000-0001-8818-0795}$^{34}$
\\
$^{1}$ SUPA Physics and Astronomy, University of St Andrews, KY16 9SS, Scotland, UK\\
$^{2}$ Department of Astronomy, University of Maryland, College Park, MD 20742-2421, USA\\
$^{3}$ Spectral Sciences Inc., 4 Fourth Ave., Burlington, MA 01803, USA\\
$^{4}$ Eureka Scientific Inc., 2452 Delmer St., Suite 100, Oakland, CA 94602, USA\\
$^{5}$ Department of Physics and Astronomy, 4129 Frederick Reines Hall, University of California, Irvine, CA 92697, USA\\
$^{6}$ Department of Physics and Astronomy, Wayne State University, 666 W. Hancock St., Detroit, MI 48201, USA\\
$^{7}$ University of Leicester, School of Physics and Astronomy, Leicester, LE1 7RH, UK\\
$^{8}$ School of Physics and Astronomy, Raymond and Beverly Sackler Faculty of Exact Sciences, Tel Aviv University, Tel Aviv 69978, Israel\\
$^{9}$ Anton Pannekoek Institute, University of Amsterdam, Postbus 94249, NL-1090 GE Amsterdam, The Netherlands\\
$^{10}$ Department of Astronomy and Astrophysics, Eberly College of Science, The Pennsylvania State University, 525 Davey Laboratory, \\ University Park, PA 16802, USA\\
$^{11}$ Institute of Gravitation \& the Cosmos, The Pennsylvania State University, University Park, PA 16802, USA\\
$^{12}$ Department of Physics, The Pennsylvania State University, University Park, PA 16802, USA\\
$^{13}$ Department of Physics, Western Michigan University, 1120 Everett Tower, Kalamazoo, MI 49008-5252, USA\\
$^{14}$ Department of Physics, Montana State University, Bozeman, MT 59717-3840, USA\\
$^{15}$ Research School of Astronomy and Astrophysics, Australian National University, Canberra, ACT 2611, Australia\\
$^{16}$ Australian Research Council (ARC) Centre of Excellence for All-sky Astrophysics (CAASTRO)\\
$^{17}$ Niels Bohr Institute, University of Copenhagen, Juliane Maries Vej 30, DK-2100 Copenhagen {\O}, Denmark\\
$^{18}$ Dipartimento di Fisica e Astronomia, Universit\`a degli Studi di Firenze, Via. G. Sansone 1, I-50019 Sesto Fiorentino (FI), Italy\\
$^{19}$ Mullard Space Science Laboratory, University College London, Holmbury St. Mary, Dorking, Surrey RH5 6NT, UK\\
$^{20}$ Astrophysics Science Division, NASA Goddard Space Flight Center, MC 661, Greenbelt, MD 20771, USA\\
$^{21}$ Joint Space-Science Institute, University of Maryland, College Park, MD 20742, USA\\
$^{22}$ Dipartimento di Fisica e Astronomia ``G. Galilei," Universit\`a di Padova, Vicolo dell'Osservatorio 3, I-35122 Padova, Italy\\
$^{23}$ INAF-Osservatorio Astronomico di Padova, Vicolo dell'Osservatorio 5 I-35122, Padova, Italy\\
$^{24}$ Department of Physics, University of Kentucky, Lexington KY 40506, USA\\
$^{25}$ Instituto de Astronom\'ia, Universidad Nacional Aut\'onoma de M\'exico. Km 103 Carretera Tijuana-Ensenada, 22860 Ensenada, B.C., M\'exico\\
$^{26}$ Department of Earth and Space Sciences, Morehead State University, 235 Martindale Drive, Morehead, KY 40351, USA\\
$^{27}$ Department of Physics and Astronomy, N283 ESC, Brigham Young University, Provo, UT 84602-4360, USA\\
$^{28}$ Space Telescope Science Institute, 3700 San Martin Dr., Baltimore, MD 21218, USA\\
$^{29}$ Department of Astronomy and Astrophysics, The Pennsylvania State University, University Park, PA 16802, USA\\
$^{30}$ Department of Astronomy, The Ohio State University, 140 W 18th Ave., Columbus, OH 43210, USA\\
$^{31}$ Center for Cosmology and AstroParticle Physics, The Ohio State University, 192 West Woodruff Avenue, Columbus, OH 43210, USA\\
$^{32}$ Astrophysics Science Division, NASA Goddard Space Flight Center, Greenbelt, MD 20771, USA\\
$^{33}$ SRON Netherlands Institute for Space Research, Sorbonnelaan 2, 3584 CA Utrecht, The Netherlands\\
$^{34}$ Department of Physics, University of California, Davis, CA 95616, USA\\
\vspace{10cm}
}
\date{Accepted XXX. Received YYY; in original form ZZZ}

\pubyear{2020}
\begin{document}
\label{firstpage}
\pagerange{\pageref{firstpage}--\pageref{lastpage}}
\maketitle
\newpage
\begin{abstract}
We present results of time-series analysis of the first year of 
the \fair\ intensive disc-reverberation campaign. We used \textit{Swift} and the Las Cumbres Observatory global telescope network to continuously monitor \fair\ from X-rays to near-infrared at a daily to sub-daily cadence. 
The cross-correlation function between bands provides evidence for a lag spectrum consistent with the $\tau\propto\lambda^{4/3}$ scaling expected for an optically thick, geometrically thin blackbody accretion disc. 
Decomposing the flux into constant and variable components,  the variable component's spectral energy distribution is slightly steeper than the standard accretion disc prediction. 
We find evidence at the Balmer edge in both the lag and flux spectra for an additional bound-free continuum contribution that may arise from reprocessing in the broad-line region.
The inferred driving light curve suggests two distinct components, a rapidly variable ($<4$ days) component arising from X-ray reprocessing, and a more slowly varying ($>100$ days) component with an opposite lag to the reverberation signal.
\end{abstract}



\section{Introduction}
Accretion discs around super-massive black holes power the most energetic persistent sources in the Universe \citep{Salpeter:1964,Zeldovich:1964,Lynden-Bell:1969}. These active galactic nuclei (AGN) allow black holes (BH) to grow throughout cosmic time \citep{Brandt:2015} and play a major role in the formation history of galaxies \citep{Fabian:2012}.
However, the combination of large distances and small angular scales precludes us from directly imaging the central engine and the vicinity of the black hole. Despite the recent progress in observing the broad-line region (BLR) via near-infrared interferometric observations \citep{Gravity:2018}, the accretion disc itself is a few orders of magnitude smaller and mostly out of reach for current instrumentation \citep[besides the notable example of M87*,][]{EventHorizon:2019}.

It is possible to use the variable nature of AGN as a mean to access information on scales well below any spatial resolution limit required to directly detect the accretion disc by imaging. By measuring the time-delayed response between the continuum light (originating in the disc) and the BLR, it is viable to infer the mass of the black hole.
This method, commonly known as reverberation mapping \citep{Blandford:1982,Peterson:2004}, has been an efficient technique to measure a sizeable sample of black hole masses spanning a broad range of luminosities and redshifts \citep[e.g.,][]{Bentz:2015,Grier:2017}.

This same basic principle can be used to study the smaller ``central engines'' of AGN.
Fluctuations in the final stage of the accretion flow modulate the flux of high-energy photons (mostly X-rays and EUV) from a hot plasma in the vicinity of the BH. These have been proposed to provide the main driver for the variability observed at ultraviolet (UV), optical and infrared wavelengths.
These high-energy fluctuations propagate outwards at the speed of light. They act as a flickering lamp-post above the BH that illuminates the surface of the accretion disc. As light travels further away from the central-most regions surrounding the BH, the high-energy fluctuations are ``echoed'' in the disc as they are absorbed and re-emitted at lower energies set by the local disc temperature.
In addition, the characteristic time delay between wavelengths/bands $[\tau(\lambda)]$ traces the average emitting radius between zones in the accretion disc allowing probe of its size and radial temperature profile \citep{Cackett:2007}. 

Continuum reverberation mapping has progressed in the last few years as a result of ambitious programmes such as the Space Telescope and Optical Reverberation Mapping Project--STORM \citep[e.g.,][]{deRosa:2015,Edelson:2015, Fausnaugh:2016,Starkey:2017} and other intensive disc-reverberation mapping (IDRM) campaigns \citep{Edelson:2017,Cackett:2018,McHardy:2018,Edelson:2019} which have found discrepancies with the standard accretion theory \citep{Shakura:1973}. 
In particular, IDRM campaigns with the Neil Gehrels \swift\ Observatory \citep[\swift\ hereafter,][]{Gehrels:2004} X-ray and UV/optical telescopes have provided continuum-lag measurements for a sample of  AGN, finding implied accretion disc sizes up to a factor of $\sim3$~times larger than predicted \citep[e.g.,][]{Edelson:2019}. 
In addition, deviations from the canonical expectations of lag as a function of wavelength ($\tau \propto \lambda^{4/3}$) in particular bands \citep[e.g., U-band:][]{Edelson:2015,Fausnaugh:2016,Cackett:2018} may point to a significant contribution of the ``diffuse continuum  emission" (DCE) from the broad-line region \citep{Korista:2001,Lawther:2018,Korista:2019,Chelouche:2019,Netzer:2020} or a different temperature profile across the disc \citep{Starkey:2017}. Furthermore, the apparent disconnect between the X-rays and UV/optical wavelengths (i.e., driver and reprocessed light respectively) may require additional reprocessing components beyond the simple accretion disc model \citep[e.g.,][]{Gardner:2017}. 

The focus of this paper is the Seyfert 1 galaxy \fair\ \citep{Fairall:1977,Ricker:1978}. It is a relatively nearby AGN \citep[$z=0.047$;][]{Hawley:1978} with a central black hole $M=2.55\pm0.56\times10^8$ \msun\ \citep{Peterson:2004}. Several X-ray studies have determined a low extinction and a lack of warm absorber \citep{Emmanoulopoulos:2011} in \fair. In addition, the persistent spectral components \citep{lohfnik:2016} suggest a clear view of the inner flow around its central engine. 
Initial variability studies found correlated X-ray and UV emission on scales $<4$~days~\citep{Lohfink:2014}. \fair\ has also been a target of continuum reverberation mapping using \swift\ \citep{Pal:2017} where inter-band lags were found to be consistent with $\tau \propto \lambda^{4/3}$, albeit with large uncertainties. 
This failure to detect a clear inter-band lag is mostly due to the low-cadence sampling ($\sim$2-7 days) and a lack of information at longer (ground-based) wavelengths in that earlier experiment.

In this paper we present the analysis and results of the first year (2018 May to 2019 February) of the IDRM campaign Key Projects\footnote{\swift\ 2018 Key Project P.I.: R.~Edelson, proposal ID: 1316154.\\ \LCO\ 2018B Key Project P.I.: R.~Edelson} with \swift\ and Las Cumbres Observatory global telescope network \citep[\LCO\ hereafter,][]{Brown:2013} on \fair. In Section~\ref{sec:observations}, we describe the multi-mission campaign as well as the data reduction and calibration. We employ the light curves to perform time-series analysis in Section~\ref{sec:timeseries} and present the wavelength-dependent lags in Section~\ref{sec:results}. Finally, we summarise our results in Section~\ref{sec:conclusions}.

\section{Observations}\label{sec:observations}

\fair\  is currently the subject of a 1.7-year IDRM campaign built around observations with \swift\ and \LCO.
This paper reports the results of roughly the first half of this campaign, covering MJD $58250 - 58550 $.
\swift\ monitored the target on an approximately daily basis in the 0.3-10 keV X-rays and six broadband UV/optical filters (UVW2, UVM2, UVW1, U, B, V) spanning $\sim 1900 - 5500 $~\AA, and \LCO\ performed \emph{uBVg$^\prime$r$^\prime$i$^\prime$z$_s$} photometry approximately three times per day (once per day at each southern site, in Chile, Australia and South Africa).
This much denser ground-based sampling aimed to compensate for a higher data loss rate, due to weather and equipment issues.
The first year of data from these two instruments form the basis of this work.
Additional monitoring was obtained, e.g., \LCO\ optical spectroscopy and {\it NICER} X-ray spectroscopy; those results will be reported elsewhere. 

\begin{figure*}
\begin{center}
	\includegraphics[width=2\columnwidth]{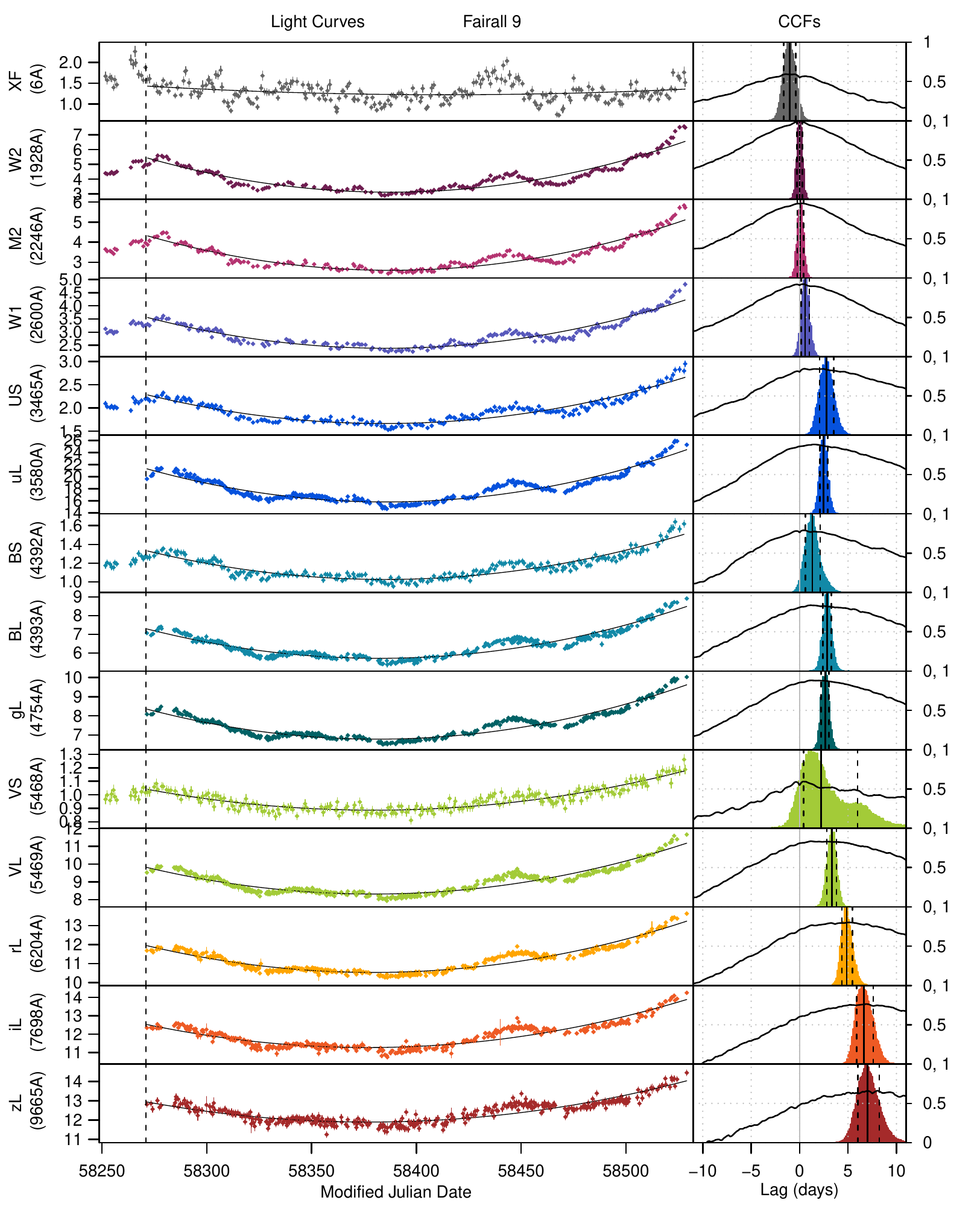}\label{fig:ccf}
    \caption{Left: \swift\ and \LCO\ light curves for \fair.
Data are ordered by wavelength, with the top panel the X-ray data from the XRT, the next three the UV light curves from the UVOT, and the remaining ten the optical light curves from both the UVOT and \LCO.
\swift\ optical UVOT data are denoted by an ``S" and the \LCO\ optical data by an ``L".
The X-ray data are in units of count~s$^{-1}$ and the optical data are all in units of mJy.
A second-order fit to the data, which was subtracted before performing the CCF analysis so as to remove excess low-frequency power, is shown as a black line.
Only data gathered after MJD 58271 were used in the CCF analysis so that the \swift\ and \LCO\ data used cover the same time period.
Right: CCFs (in black; scale on the right) and FR/RSS centroid distributions (in colour) for each band relative to the W2 band.
A positive value means the comparison band lags behind W2.
The median of the distribution and its 1$\sigma$ confidence intervals are shown as black solid and dashed lines, respectively. These values are presented in Table~\ref{tab:log_observations}.
}
\end{center}
\end{figure*}

\subsection{Swift - XRT}
Data from two of \swift's three instruments are used in this paper: the X-Ray Telescope \citep[XRT,][]{Burrows:2005} and the UltraViolet/Optical Telescope \citep[UVOT,][]{Roming:2005}.
All XRT observations were made in photon counting (PC) mode.
The XRT data are analysed using the standard \swift\ analysis tools described by \cite{Evans:2009}\footnote{\url{http://www.swift.ac.uk/user\_objects}}.
These produce light curves that are fully corrected for instrumental effects such as pile up, dead regions on the CCD, and vignetting.
The source aperture varies dynamically according to the source brightness and position on the detector. 

We output the observation times (the midpoint between the start and end times) in MJD for ease of comparison with the UVOT data.
We utilise ``snapshot'' binning, which produces one bin for each continuous spacecraft pointing. 
This is done because these short visits always occur completely within one orbit with one set of corresponding exposures in the UVOT filters.
In all other cases we used the default values.
We generated an X-ray light curve covering the full XRT bandpass (XF; 0.3--10~keV).
For a detailed discussion of this tool and the default parameter values, please see \cite{Evans:2009}.

\subsection{Swift - UVOT}\label{sec:uvot}

This paper's UVOT data reduction follows the same general procedure described in our previous work \citep[e.g.,][]{Edelson:2019}.
This process has three steps: flux measurement, removal of points that fail quality checks, and identification and masking of low sensitivity regions of the detector.
Each step is described in turn below.

All data were reprocessed for uniformity (using version 6.22.1 of {\tt HEASOFT}) and their astrometry refined \citep[following the procedure of][]{Edelson:2015} before measuring fluxes using {\tt UVOTSOURCE} from the {\tt FTOOLS}\footnote{\url{http://heasarc.gsfc.nasa.gov/ftools/}} package \citep{Blackburn:1995}.
The filters and other details of this instrument are given in \cite{Poole:2008,Breeveld:2011}.
Source photometry was measured in a circular extraction region of 5\arcsec\ in radius, while backgrounds were taken from concentric 40\arcsec--90\arcsec\ annuli.
In the {\it V} band in particular, especially when the AGN power is lower, the galaxy contributes significantly to the measured flux.
The final flux values include corrections for aperture losses, coincidence losses, large-scale variations in the detector sensitivity across the image plane, and declining sensitivity of the instrument over time.

In the second step, the resulting measurements are used for both automated quality checks and to flag individual observations for manual inspection. 
These automated checks include aperture ratio screenings to catch instances of extended point-spread functions (PSFs) or when the astrometric solution is off, and a minimum exposure time threshold of 20~s.
Data are flagged for inspection when the fitted PSFs of either the AGN or several field stars were found to be unusually large or asymmetric, or if fewer than 10 field stars with robust centroid positions are available for astrometric refinement.  
Upon inspection, observations are rejected if there are obvious astrometric errors, doubled or distorted PSFs, or prominent image artifacts (e.g., readout streaks or scattered light) that would affect the AGN measurement.
Note that we adopt a non-standard setting of 7.5\% for the {\tt UVOTSOURCE} parameter {\tt FWHMSIG} because this yields flux uncertainties more consistent with Gaussian statistics \citep{Edelson:2017}.

The third step screens out data that fall within detector regions with reduced sensitivity.  
In our previous work, these regions were identified through a bootstrap method based on the prominent low outliers from the AGN light curves  \citep[e.g.,][]{Edelson:2019}. We have now mapped the sensitivity variations with more uniform, higher-resolution coverage across the entire detector creating sensitivity masks for each filter, based on an independent set of data. 
This process is described in Appendix~\ref{sec:app_map}.
The light curves that result after applying these masks are shown in Fig.~\ref{fig:ccf}.

\subsection{Las Cumbres Observatory}
\label{sec:lco_obs}

We obtained a multi-colour light curve by using the \LCO\ network's 1-m robotic telescopes to monitor \fair, as part of the 2018B AGN Key project. Our observations were obtained with an average cadence of three observations per day, one at each of the southern \LCO\ sites located at Sutherland (South Africa), Cerro Tololo (Chile), and Siding Spring (Australia). 
Each observation consisted of two consecutive exposures with the Sinistro CCD camera, to mitigate the impact of cosmic ray hits and to provide an internal consistency check on the uncertainties.
In total, 6109 individual exposures were taken in seven bands: \emph{B}, \emph{V}, \emph{$u^\prime$}, \textit{$g^\prime$}, \emph{$r^\prime$}, \emph{$i^\prime$} and \emph{z$_s$}. A summary of the observations is presented in Table~\ref{tab:log_observations}.

The data extracted from the \LCO\ archive were bias and flat-field corrected images pipeline processed with {\sc banzai} \citep{curtis_mccully_2018_1257560}.
We extracted multi-aperture aperture photometry with {\sc Sextractor} \citep{bertin:1996} on every image. We constructed a global background model by smoothing the image in a 200 pixel mesh, large enough to avoid the extended sources influencing the background estimate. After subtracting the background model and performing the aperture photometry, we constructed a curve of growth for each individual frame and measured the correction factor required to bring every star flux as if extracted from an azimuthally averaged point-spread function (PSF). This method produced stable light curves which were more robust to the diverse range of atmospheric conditions (e.g., airmass, seeing) taken throughout the year, without performing PSF extraction.
The colour-correction and atmospheric extinction coefficients were obtained from \citet{valenti:2016} and applied before the photometric calibration.
We used comparison stars in each field to perform an image zero-point calibration at each epoch. We used the \textit{AAVSO Photometric All-Sky Survey} (APASS) DR10 \citep{henden:2018} for all \LCO\ filters except the \textit{u}~band, for which no APASS information was available. Here, we made use of the \textit{Swift}/UVOT U band images to obtain the fluxes of reference stars in the field. We applied a 3-$\sigma$ clipping to the zero-point estimates and perform a bootstrap method to obtain the error on the zero-point. The full light curves for year 1 are shown in Fig.~\ref{fig:ccf} and the data format is shown in Appendix~\ref{sec:app_data}.

After these corrections, we noted small but significant systematic offsets among the light curves from different sites and telescopes in the \LCO\ network, in particular for the \textit{B}~band (see Fig.~\ref{fig:fig2}), but evident also at \textit{g'} and \textit{z$_s$}. 
We therefore performed an inter-telescope calibration, as described below, which reduced the offsets to obtain the light curves shown in Fig.~\ref{fig:ccf}. 

\begin{table*}
	\centering
	\caption{Observation log and light curve properties of the first year of \fair\ campaign with \swift\ and \LCO.}
	\label{tab:log_observations}
	\begin{tabular}{lcccccccc|cc|cc}
	\hline
	Filter & $\lambda_{\rm eff}$ & $t_{\rm exp}$ & Epochs&$\Delta t_{\rm mean}$& $\Delta t_{\rm med}$& $\left<F_\nu\right>$ & $F_\nu$ Host& $F_{\rm var}$ &$r_{\rm max}$&$\tau_{\rm CCF}$ &$1\sigma$ &$\tau_{\rm {\sc cream}}$\\
		& \AA & s & & day & day& mJy & mJy&  & &day & (16\%,84\%) & day \\
	&(1)&(2)&(3)&(4)&(5)&(6)&(7)&(8)&(9)&(10)&(11)&(12)\\
	\hline
	Swift &  &  & & & & & & & &CCF&CCF & {\sc cream}\\
	\hline
	XF   &    6 & 1000 & 263& 1.074 & 0.997 & $\cdots$ &.  $\cdots$&$0.184\pm0.004$.        & 0.594  &  -1.02  & (-1.63,-0.39)\\
	W2 & 1928 & 333 & 240& 1.184 & 0.998 & $5.93\pm 0.08$ & $0.12\pm0.12$  &$0.227\pm0.001$ & 1.0    &   0.0   & (-0.27, 0.31)  & $0.00\pm 0.10$\\
	M2 & 2246 &  250 & 235& 1.209 & 1.006 & $5.98\pm 0.24$ & $0.61\pm0.13$  &$0.204\pm0.001$& 0.95   &   0.08  & (-0.25, 0.40)  & $0.25\pm 0.11$ \\
	W1 & 2600 &  167 & 232& 1.225 & 1.048 & $5.86\pm 0.23$ & $1.65\pm0.12$  &$0.166\pm0.001$& 0.922  &   0.58  & ( 0.17, 1.00)  & $0.55\pm 0.13$ \\
	U    & 3465 & 83 & 247& 1.150 & 0.999 & $5.72\pm 0.22$ & $3.04\pm0.12$  &$0.134\pm0.002$& 0.846  &   2.77  & ( 2.06, 3.53)& $1.33\pm 0.17$   \\
	B    & 4392 & 83 & 255& 1.114 & 0.998 & $4.34\pm 0.20$ & $3.90\pm0.09$  &$0.107\pm0.002$& 0.791  &   1.3   & ( 0.60, 2.12)& $2.26\pm 0.23$   \\
	V    & 5468 & 83 & 247& 1.150 & 1.001 & $3.99\pm 0.23$ & $6.26\pm0.08$  &$0.076\pm0.002$& 0.595  &   2.22  & ( 0.43, 6.00)& $3.35\pm 0.28$  \\
	\hline
	\LCO\ &  &  & &\\
	\hline
	\emph{$u^\prime$}   & 3580 &300 & 874 & 0.357 & 0.121 &$5.42\pm0.18$ & $2.87\pm0.11$ &$0.117\pm0.002 $& 0.883 &  2.47 & ( 2.07, 2.92) &$1.33\pm0.17$  \\
	\emph{B}            & 4392 & 60 & 914 & 0.322 & 0.039 &$4.35\pm0.13$ & $2.92\pm0.09$ &$0.105\pm0.002 $& 0.837 &  2.85 & ( 2.40, 3.29) &$2.26\pm0.23$  \\
	\emph{$g^\prime$}   & 4770 & 60 & 861 & 0.355 & 0.059 &$4.43\pm0.13$ & $3.94\pm0.09$ &$0.092\pm0.001$ & 0.881 &  2.64 & ( 2.23, 3.04)  &$2.64\pm0.25$ \\
	\emph{V}      & 5468 & 60 & 850 & 0.411 & 0.077 &$4.48\pm0.14$ & $5.42\pm0.09$ &$0.077\pm0.001$ & 0.836 &  3.32 & ( 2.82, 3.80)  &$3.35\pm0.28$ \\
	\emph{$r^\prime$}   & 6215 & 60 & 857 & 0.432 & 0.072 &$4.17\pm0.13$ & $7.84\pm0.09$ &$0.058\pm0.001$ & 0.802 &  4.88 & ( 4.36, 5.46)  &$4.08\pm0.30$ \\
	\emph{$i^\prime$}   & 7545 & 60 & 880 & 0.419 & 0.064 &$4.01\pm0.14$ & $8.68\pm0.08$ &$0.055\pm0.001$ & 0.764 &  6.66 & ( 5.88, 7.59)  &$5.26\pm0.30$ \\
	\emph{z$_s$}& 8700 &120 & 828 & 0.360 & 0.055 &$3.32\pm0.13$ & $9.75\pm0.07$ &$0.044\pm0.001 $& 0.662 &  7.03 & ( 6.02, 8.22) &$6.12\pm0.29$  \\
	\hline
	\end{tabular}
{\raggedright {\sc Note--} (1)$\lambda_{eff}$ is the effective wavelength at each band. (2) $t_{exp}$ is the individual exposure time of each frame. The \swift\ exposure times are the average for a 1 ks exposure for \swift\ UVOT mode 0x30ed. (3) Epochs are the number of independent measurements. (4) Mean and (5) median time between consecutive images  in each filter. (6) $\left<F_\nu\right>$ is the average flux density of the variable component and (7) $\left<F_\nu\right>$ Host, for the underlying host galaxy.
(8) $F_{\rm var}$ is the fractional variability as described in \citet{Fausnaugh:2016}. (9) $r_{\rm max}$ is the maximum correlation coefficient from the CCF analysis. (10) $\tau_{\rm CCF}$ is the median value in the inter-band lag distribution its $1\sigma$ 16\% and 84\% confidence interval (11). (12) $\tau_{\rm {\sc cream}}$ are the mean values of the delay distributions as measured with {\sc cream}. All lag measurements for CCF and {\sc cream} are shown in reference to W2 band.
\par}
\end{table*}

\subsubsection{Intercalibration of the \LCO\ light curves}\label{sec:intercal}

\begin{figure}
\includegraphics[trim={0cm 0.5cm 1cm .2cm}, clip,width=1.\columnwidth]{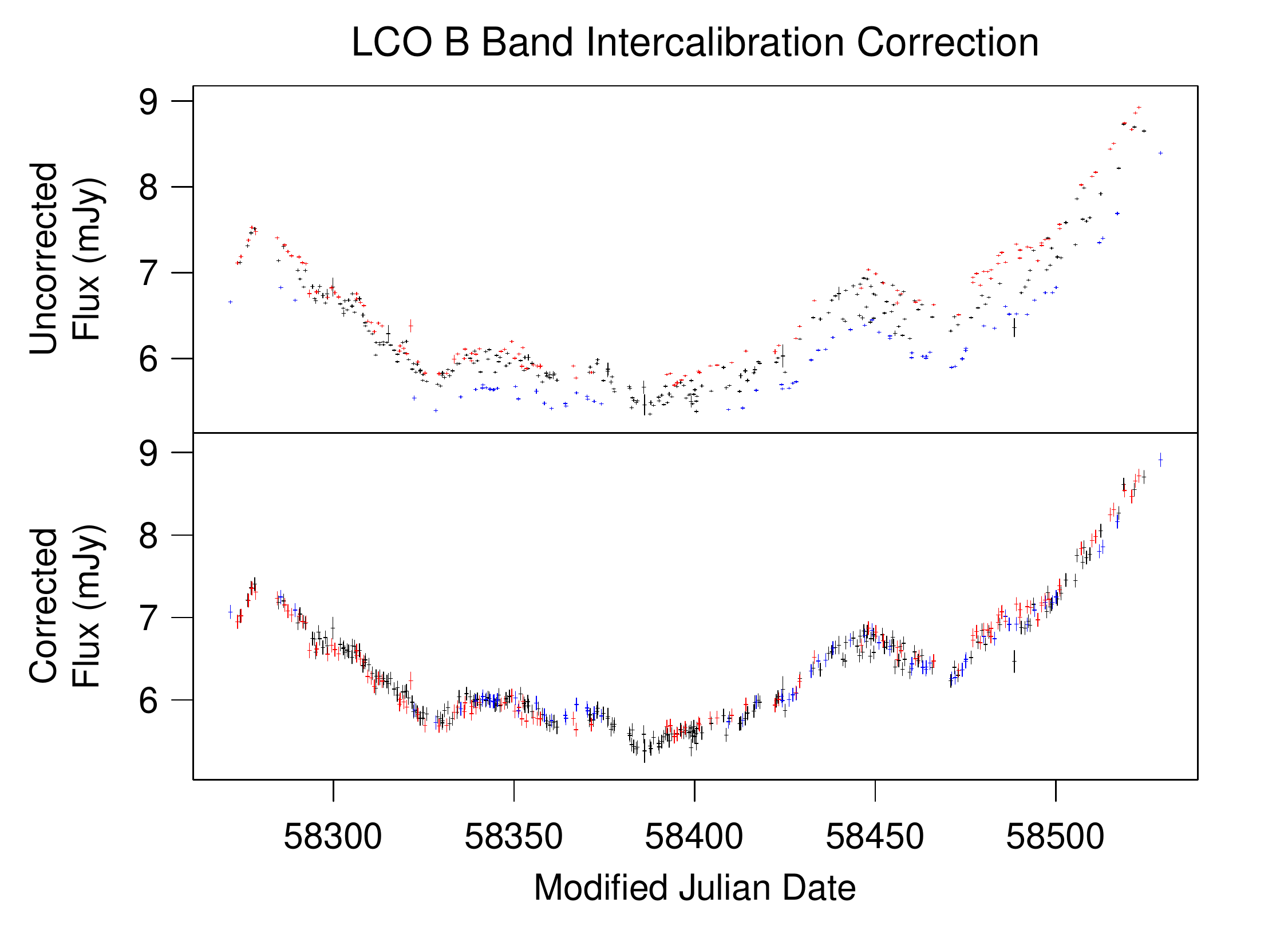} 
\caption{\LCO\ $B$-band light curve of \fair\ showing the data before and after the inter-telescope flux recalibration.
Eight of the \LCO\ southern telescopes were used to monitor \fair, at sites in Chile, South Africa and Australia.
Data from the Chilean ``1m004'' and ``1m005'' LCO
telescopes are shown in blue and red respectively, while those for the other six telescopes (in all three sites) are shown in black.
Note that in the upper panel the (uncorrected)  red and blue light curves differ by many times the errors, while in the lower panel the (corrected) data appear consistent within the final errors.
The recalibration process is discussed in Section~\ref{sec:intercal}.
}
\label{fig:fig2}
\end{figure}

A total of eight \LCO\ 1-m telescopes on three southern continents were used in this campaign. 
Despite the fact that identical designs were used for the telescopes, detectors and filter sets, we detected small systematic offsets, as shown in Fig.~\ref{fig:fig2}.
Thus we found it necessary to recalibrate the light curves we extracted from the CCD images taken by these different telescopes.
This was done in a 3-step process.
First the data are separated by band with unique identifiers noted for each of these eight telescopes.
Then for each data point, the ratio of the reported flux to that of all other fluxes gathered (a) within 2 days and (b) on different telescopes is determined.
Then for each unique band/telescope combination, the mean of this ratio is determined and all fluxes and errors divided by this mean amount.  
This yields better-intercalibrated fluxes with smaller dispersion around the mean value in all bands.

Thus the second step is to iterate this process, a total of six times, until the changes in this calibration factor become negligible.
The final result is effectively a first-order (multiplicative) intercalibration of the telescopes. 

This strategy corrects the measured fluxes but does not affect the reported uncertainties, which are still small compared to the short term dispersion within these data.
That is, systematic calibration errors are still present and must be quantified.
Thus the third step adjusted the reported uncertainties by determining an additional ``inter-telescope calibration error'' term to be added in quadrature to the reported uncertainties.
This is done by first taking the previously calibrated flux values in each band (no longer segregating by telescope) and measuring the absolute values of the point-to-point differences between successive points, i.e., the moving range $MR_i=|x_i - x_{i-1}|$.
We then use the factors method to estimate of the total short-term dispersion $\sigma_{dis} = d_2 \cdot \langle MR_i\rangle$, where the $d_2=1.128$ is the unbias factor\footnote{$d_2=1.128$ is the expected value of the range of two observations from a normal population with standard deviation = 1.} \citep{Wheeler:1992}.  
We assume that intrinsic variations are small compared to the errors on the data on these timescales as the average sampling is $\sim$0.7 day in this well-sampled data set.
Then we determine the additional variance that, when added to the sum of the squares of the observed errors, yields the total variance implied by the short-term dispersion. Values for each telescope and filter are shown in the Appendix~\ref{sec:app_data} and Table~\ref{tab:uncertainties}.
The amplitude of this additional error term (the square root of the additional variance) ranges from 0.58\% of the mean flux in the best band (\textit{r}) to 1.23\% in the worst bands (\textit{B} and \textit{z}).
In all cases it is much larger, typically by a factor of 3-6, than the uncorrected errors.
Thus we conclude that inter-telescope calibration is the dominant source of error in our \LCO\ light curves.

\section{Time series analysis}\label{sec:timeseries}
\subsection{Cross-correlation function}\label{sec:ccf}

Cross-Correlation Functions (CCFs) were measured using the interpolated cross-correlation function \citep{Gaskell:1987}.
We used the {\tt sour} code\footnote{This code is available at \url{https://github.com/svdataman/sour}}, which is based on the specific implementation presented in \cite{Peterson:2004}.
We first normalised the data by subtracting the mean and dividing by the standard deviation.
These were derived ``locally'' --- only the portions of the light curves that are overlapping in time for a given lag are used to compute these quantities.
As discussed by \cite{White:1994}, ``global'' normalisation using the entire light curve is appropriate for stationary variability processes but local normalisation is more appropriate for weakly non-stationary variability, as has been observed to be the case in AGN.
Due to the uneven sampling of the light curves, we require to perform an interpolation to properly perform any CCF analysis.
We implemented ``2-way'' interpolation, which means that for each pair of bands we first interpolated in the ``reference'' band and then measured the correlation function, next interpolated in the ``subsidiary'' band and measured the correlation, and subsequently averaged the two to produce the final CCF.
This was done because the alternative ``1-way'' interpolation produces an autocorrelation function that is not strictly symmetric, which, as an even function, it should be.
The W2 light curve is always the reference and the other bands are considered the subsidiary bands in this analysis.
This band was chosen because it has the shortest UV wavelength and thus is closest to the thermal peak of the accretion disc.
The CCF ($r(\tau)$ where $\tau$ is the lag) is then measured and presented to the right of the light curves in Figure~\ref{fig:ccf}.

We then used the ``flux randomisation/random subset selection'' (FR/RSS) method \citep{Peterson:1998} to estimate uncertainties on the measured lags.
This is a Monte Carlo technique in which lags are measured from multiple realisations of the CCF.
The FR aspect of this technique perturbs in a given realisation each flux point consistent with the quoted uncertainties assuming a Gaussian distribution of errors.
In addition, for a time series with $N$ data points, the RSS randomly draws with replacement $N$ points from the time series to create a new time series.
In that new time series, the data points selected more than once have their error bars decreased by a factor of $n_\mathrm{rep}^{-1/2}$, where  $n_\mathrm{rep}$ is the number of repeated points. 
Typically a fraction of $\left(1-{\frac{1}n}\right)^n\rightarrow 1/e$ of data points are not selected for each RSS realisation. 
In this paper, the FR/RSS is applied to both the ``reference'' and subsidiary light curves in each CCF pair.
The CCF ($[r(\tau)]$ where $\tau$ is the lag) is then measured and a lag determined to be the weighted mean of all points with $ r > 0.8\, r_\mathrm{max} $, where $ r_\mathrm{max} $ is the maximum value obtained for the correlation coefficient $r$ \citep[e.g.,][]{Edelson:2019}. Values for every band are given in Table~\ref{tab:log_observations}.
For the data presented herein, lags are determined for 55,000 realisations and then used to derive the median centroid lag and 68\% confidence intervals, also presented in Table~\ref{tab:log_observations}.
This number of trials was chosen so that uncertainties on the derived median lags and confidence intervals due to sample statistics are negligible.
Repeating this test confirms that these quantities change by only very small amounts compared to the widths of the confidence intervals.

\subsection{CREAM light curve modelling}\label{sec:cream}

Following \citet{Starkey:2016,Starkey:2017}, we fit the \swift\ and \LCO\ light curves using the {\sc cream}\footnote{The {\sc python} implementation of this software {\sc pycecream} can be found at \url{https://github.com/dstarkey23/pycecream}} reverberating accretion disc model in order to retrieve physical parameters from the system.
Here, we briefly discuss the reverberation methodology. {\sc cream} models the delay distribution of the accretion disc at every wavelength assuming local reprocessing of the high-energy photons and re-emission as a blackbody spectrum. For the standard accretion disc  temperature profile, $T^4\propto M \, \dot{M} \, R^{-3}$, the mean time delay scales as  $\langle\tau\rangle \propto \left(M\, \dot{M}\right)^{1/3} \lambda^{4/3}$. Thus, by using multi-wavelength data, we can fit for the product $M \dot{M}$, given by the average delay of the individual light curves. Since the lamp-post model assumes that the same compact source is responsible for driving the variations in all of the echo light curves,
it is necessary to define the driving light curve. {\sc cream} builds an independent model for this driving light curve (as opposed to other methods that require one to be provided) using a damped-random walk (DRW) model as a prior. Each echo light curve is then modelled at their corresponding wavelengths by convolving the driving light curve with the best-fit delay distribution, and then scaled by a multiplicative factor and shifted by an additive constant. We explore the parameter space using a Markov Chain Monte Carlo (MCMC) procedure to find the posterior distributions of the parameters of interest.

We have used a face-on disc \citep[$i=0$, the effect of inclination angle on the measurement of the delay spectrum is very small, as shown in][]{Starkey:2016} and a standard temperature profile $T\propto R^{-3/4}$. The noise model contains an extra scatter term added in quadrature for each light curve. The driving light curve has a fixed maximum Fourier frequency of $\nu_{\rm max}=2\pi/{\Delta t_{\rm mean}} = 1.0$ cycles/day, set by the average time separation between observations in the \LCO\ bands (see Table~\ref{tab:log_observations}).
We ran the MCMC procedure for $10^5$ iterations where we judged that the chains had adequately sampled the posterior distributions of the model parameters. We discarded the first 20,000 chains as burn-in.

\begin{figure}
	\includegraphics[width=1.\columnwidth]{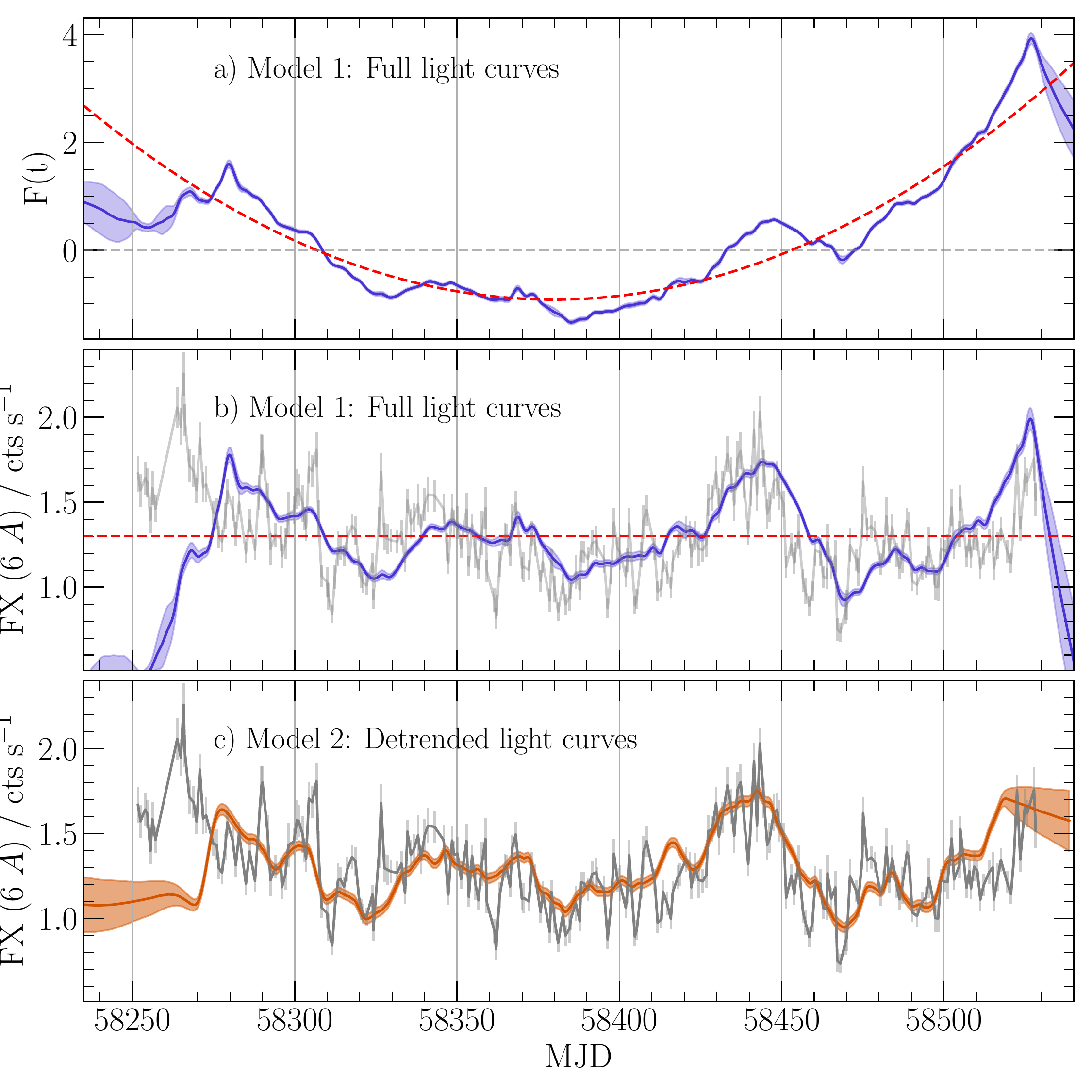}
    \caption{\textit{a)} Driving light curve derived by {\sc cream} with its 68\% confidence interval as the contours for the fit to the full light curves. A  quadratic fit used to detrend is shown as the red dashed line. \textit{b)}  Comparison between the full \swift\ X-ray 0.3-10 keV light curve and the driving light curve after subtracting a quadratic trend. The driving light curve has been scaled to match the X-ray one. A clear shift between these two light curves is observed. \textit{c)} Driving light curve obtained from the fitting of the detrended light curves with {\sc cream} compared to X-rays.}
    \label{fig:driving_lc}
\end{figure}

We made an initial fit (Model 1) using the full light curves with no detrending. 
This resulted in delay distributions with mean lags that were shorter by a factor of $\sim2-3$ than those obtained via CCF. 
Since {\sc cream} generates the driving light curve instead of using one as a reference (as in the case of CCF), we can compare it directly to the X-ray light curve to test whether the latter is the main driver of the reprocessing. In the first panel of Fig.~\ref{fig:driving_lc}, we show the driving light curve which presents a long-term variation that is not present in the X-ray one (shown in the second panel), and most likely not arising from the reprocessing itself.  However, variability superimposed on this slow and smooth trend is visible, at shorter time-scales (of order days) compatible with a reverberation signal. To test this, we fitted a quadratic polynomial to the driving light curve to focus on the shorter time-scale, shown in the second panel of Fig.~\ref{fig:driving_lc}. The residual variability now resembles qualitatively the 0.3-10 keV X-ray light curve, albeit with a discernible lag between them. This particularly notable in the bump around MJD 58440. 

In order for the delay distribution to be modified by this slow component, it must carry an intrinsic time-dependency in addition to the dilution of the amplitude of the signal. We discuss this component in further detail in Section~\ref{sec:slow}. Motivated by this result, we performed a second fit with {\sc cream} (Model 2) with the same parameters and priors described earlier after subtracting the quadratic detrending i.e., using the same signal as those used by the CCF analysis. The resulting fit to each light curve, residuals of the fit, and the delay distributions for each band are shown in Appendix~\ref{sec:app_cream}, Fig.~\ref{fig:cream_fit1} and \ref{fig:cream_fit2}. By removing this slow component, we recovered a lag spectrum with values that match closely to those obtained by CCF, shown in reference to the UVW2 band in Table~\ref{tab:log_observations}. We note that the largest disagreements between CCF and {\sc cream} are in the $U$ and $i$ band, likely arising from the contribution of the diffuse continuum emission (see Section ~\ref{sec:dce} for further discussion) that currently {\sc cream} does not incorporate and only assumes the disc $\tau\propto\lambda^{4/3}$ lag distribution.
Furthermore, the driving light curve inferred now directly resembles the X-ray light curve shown in the two bottom panels of Fig.~\ref{fig:driving_lc}. The mismatch between the driving light curve and the X-rays at both ends is caused by the inadequate representation of the smooth component as a parabola. Since this component flattens out during, the parabola overpredicts the contribution thus resulting in large residuals.

This indicates that there are (at least) two processes that contribute to the overall variability of \fair\ and that the observed X-rays cannot be the sole driving mechanism of variability at optical wavelengths (see further discussion in Section~\ref{sec:slow}). In order to address the origin of such long-term variability, we require a larger time baseline. Thus, continuing the multi-year monitoring of \fair\ is paramount.

The {\sc cream} fit estimates the posterior probability distribution of the product
$M\,\dot{M}$. We find a very well determined value of $\log( M\,\dot{M} /\, {\rm M}_{\odot}^2\, {\rm yr}^{-1} ) =7.49\pm0.07$.
Assuming a $2.55\times10^8$~\msun\ black hole \citep{Peterson:2004}, the Eddington ratio is therefore $\dot{m}_{Edd}=0.020\pm0.004$, consistent with the value determined from X-ray studies \citep{Vasudevan:2009}. 

\section{Results}\label{sec:results}
\subsection{Interband Lag fits}

\begin{figure}
	\includegraphics[width=1.\columnwidth]{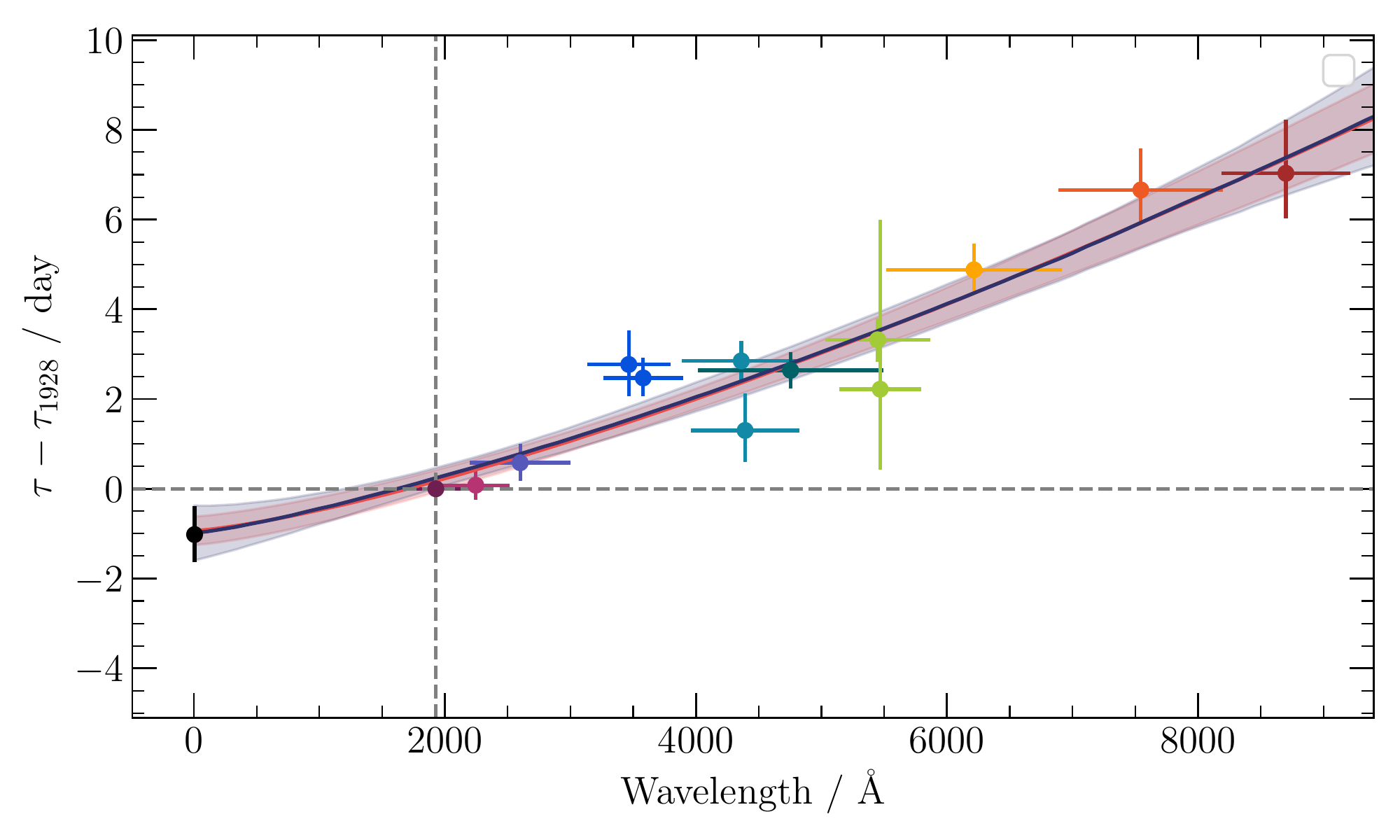}
	\includegraphics[width=1.\columnwidth]{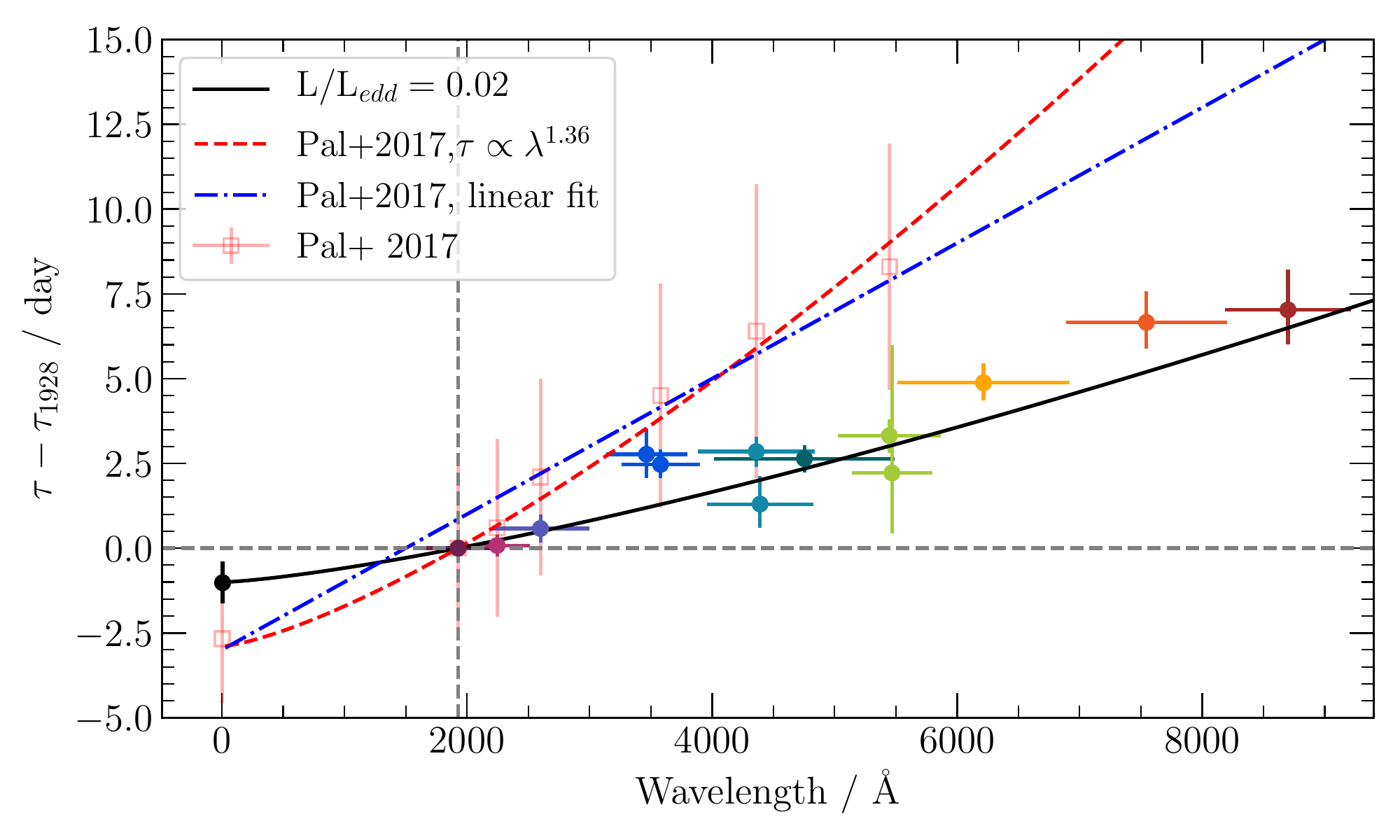}
    \caption{Inter-band lag spectrum of \fair. \textit{Top:} CCF lags as a function of wavelength for the \swift\ and \LCO\ bands. All lags are measured with respect to the \swift/UVW2 band. The best fit relation for the reprocessing model ($\tau\propto\lambda^{4/3}$) is shown as the red line and its $1\sigma$ error envelope is the shaded region. The black solid line and error envelope shows the fit when leaving the power-law index as a free parameter. \textit{Bottom:} Lag spectrum comparison between this work and the inter-band lags measured by \citet{Pal:2017}. We show their linear (blue) and exponential fits (red). The predicted lag spectrum for \fair\ is shown as the black line. This shows the clear need of IDRM campaigns to extract the lag spectrum with high accuracy.}
    \label{fig:lag_vs_wavelength}
\end{figure}

The standard model for lamp-post irradiation of a thin accretion disc predicts a lag-wavelength dependency of $\tau\propto\lambda^{4/3}$ \citep{Cackett:2007}. The high quality of the CCF lags detected in Sec.~\ref{sec:ccf} allows us to test this prediction. We fitted the CCF lag spectrum, $\tau$, as a function of wavelength, using the following functional form:
\begin{equation}\label{eq:lag_spec}
    \tau = \tau_0 \, \left[\left(\frac{\lambda}{\lambda_0}\right)^{\alpha} - y_0\right],
\end{equation}
where $\tau_0$ is the amplitude the power-law lag spectrum,  $\alpha$ is the power-law index, $\lambda_0=1928$~\AA\ is the reference wavelength (UVW2 filter) and $y_0$ allows the model lag at $\lambda_0$ to differ from 0.  Our fit uses only the UV and optical lags,
omitting the X-ray/UVW2 lag.

The CCF lag distributions are close to Gaussian, but with slight asymmetries. We averaged the low and high 1$\sigma$ confidence intervals to use as our 1$\sigma$ standard deviation. As with previous studies, we first fitted our delay spectrum to Eq.~\ref{eq:lag_spec} for a fixed power-law index of $\alpha=4/3$ and then fit with $\alpha$ as a free parameter, omitting the X-rays. In both cases, we find a consistent agreement with the standard accretion disc scaling prediction ($\tau\propto\lambda^{4/3}$), with a reduced
$\chi^2/11=1.16$. In addition, when fitting for the slope, we find $\alpha=0.93\pm 0.39$,
closer to a linear relation, with a reduced $\chi^2/10=1.17$. The full list of best-fit parameters is shown in Table~\ref{tab:lag_fit}.

The high-quality X-ray light curve (and thus lag detection) and its clear connection between the optical and UV bands (see Section~\ref{sec:cream}) allows us to better constrain the lag spectrum. In Fig.~\ref{fig:lag_vs_wavelength} we show the best fit to the CCF lags for a fixed power-law index of $\alpha=4/3$ and then fit with $\alpha$ as a free parameter. Both fits overlap and produce consistent results for the size of the disc.
\begin{table*}
	\centering
	\caption{Best fit parameters to the inter-band lag spectrum}
	\label{tab:lag_fit}
	\begin{tabular}{ccc|cc|cc}
	\hline
     & \multicolumn{2}{c}{No X-rays}& \multicolumn{2}{c}{Including X-rays}&
     \multicolumn{2}{c}{Including X-rays \& no U-band}\\
     Parameter & $\alpha$ Fixed & $\alpha$ Free& $\alpha$ Fixed & $\alpha$ Free&
     $\alpha$ Fixed & $\alpha$ Free\\
     \hline
     $\tau_0/$day & $1.20\pm0.10$ & $2.32\pm1.65$ & $1.20\pm0.09$& $1.40\pm0.48$ & $1.19\pm0.08$ &
     $1.32\pm0.46$\\
     $\alpha$ & 4/3 & $0.93\pm0.39$ & 4/3 & $1.22\pm0.22$&4/3 & $1.26\pm0.23$\\
     $y_0$ &$0.92\pm0.16$ &$1.02\pm0.09$ & $0.91\pm0.15$& $0.93\pm0.13$&
     $0.94\pm0.15$& $0.95\pm0.15$\\
     \hline
     $\chi^2_{\nu}$ & 1.16 &1.17 & 1.09& 1.175& 1.017 & 1.118\\
    dof & 11 & 10 & 12 & 11 & 10 & 9\\
     \hline
\end{tabular}
\end{table*}

Previous studies have found CCF lag spectra suggesting that disc sizes are larger than expected by factors $\sim2-3$. This \textit{``too-big disc" problem} has been noted in a number of systems so far e.g., \citet{Edelson:2015,Fausnaugh:2016,Edelson:2017,Cackett:2018,Pozo-Nunez:2019,Edelson:2019}. 
In Fig.~\ref{fig:lag_vs_wavelength}, we show the lag spectrum prediction following \citet{Fausnaugh:2016} and  \citet{Pal:2017}:

\begin{equation}
    \tau - \tau_0 = \frac{1}{c} \left(\frac{\lambda_0}{k}\right)^{4/3}\left(\frac{3\,G\,M\,\dot{M}}{8\,\pi\,\sigma} + \frac{(1-A)\,L_x\,H}{4\, \pi\,\sigma}\right)^{1/3}\left[\left(\frac{\lambda}{\lambda_0}\right)^{4/3} - 1\right]
\end{equation}
where $k = 2.897 \times 10^{-3}$ mK is Wien's constant, $\sigma$ is the Stefan-Boltzmann constant, $A=0.2$ is the albedo and $H=6\,G\,M/c^2$ is the height of the illuminating X-ray source. We have used an Eddington ratio $L/L_{\rm Edd}=0.02$ \citep{Vasudevan:2009}, to provide a direct comparison with the previous study of this source. We find a $\chi^2_\nu = 13.049 / 12 = 1.087$ for the accretion disc prediction (omitting the U-band and including the X-rays).
In \citet{Pal:2017}, the observed lags using only \swift/UVOT provided evidence for a larger disc as shown in the bottom panel of Fig.~\ref{fig:lag_vs_wavelength} albeit a lower signal-to-noise ratio of their lag measurements. 
We note the strong disagreement between the accretion disc prediction and their lags measured, especially at long wavelengths.
This demonstrates that IDRM, especially including long-wavelength ground-based measurements, can sharpen the tests of accretion disc theory predictions.

\subsection{Spectral energy distribution}

We calculated the fractional variability ($F_{\rm var}$) as a function of wavelength \citep{Rodriguez:1997,Vaughan:2003}. We followed the procedure described by \citet{Fausnaugh:2016} and the results for each light curve are shown in Table~\ref{tab:log_observations}. We find a decreasing value of $F_{\rm var}$ as a function of wavelength in the UV/optical bands. This is likely due at least in part to the contribution of the underlying galaxy, which has a redder spectral energy distribution (SED) than the variable AGN and will dilute the variability at longer wavelengths. Therefore, we expect that the shorter wavelengths trace closely the intrinsic variability of the AGN.

\begin{figure*}
	\includegraphics[width=16cm]{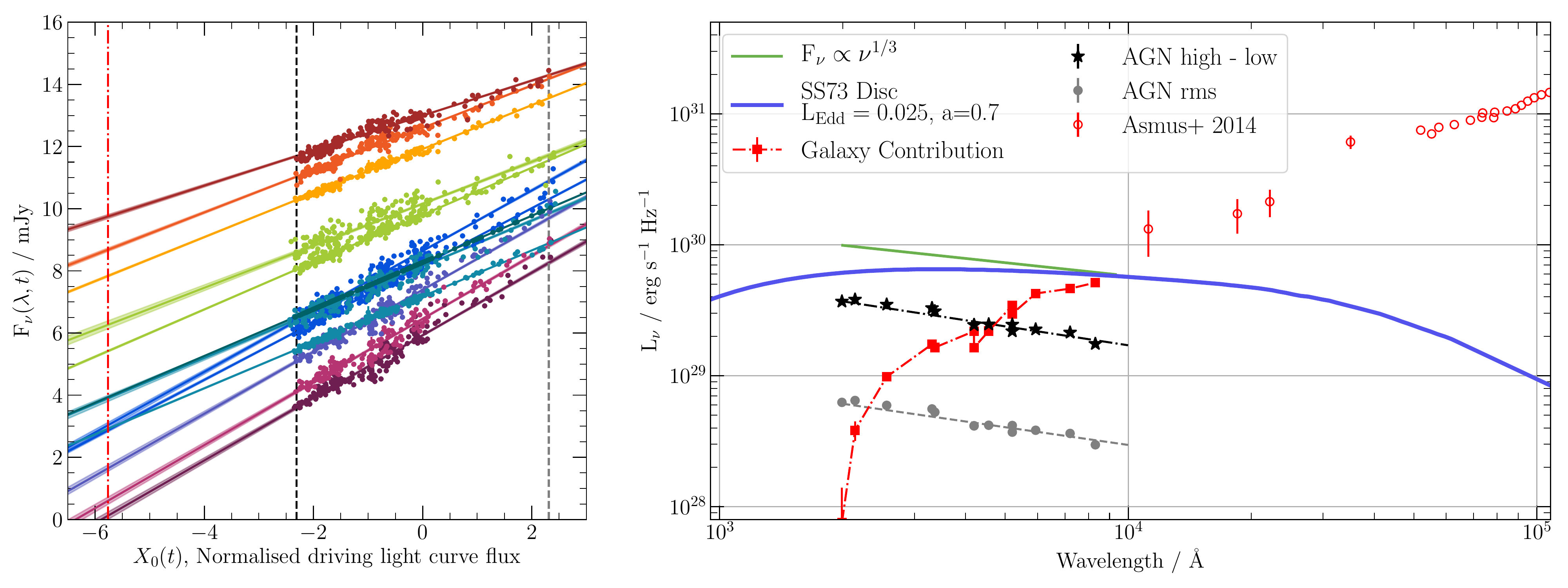}
    \caption{\textit{Left:} Spectral decomposition of the variable component of \fair\ for all \swift\ and \LCO\ light curves. The flux-flux plot uses the driving light curve obtained with {\sc cream}. Vertical lines show the underlying galaxy contribution (red) as well as the AGN high (grey) and low (black) variable spectra. \textit{Right:} Spectral energy distribution of the high-low (\textit{black}) and RMS (\textit{grey}) variable component -- AGN -- as well as the constant component --galaxy+DCE (red). The mid- and far-infrared measurements are shown as the open squares, taken from \citet{Asmus:2014MNRAS}. All SEDs are corrected for extinction, see text for details. The variable component are compared against a standard disc SED with a $L_{\rm Edd}=0.02$ and a BH spin, $a=0.7$.}
    \label{fig:galaxy_spec}
\end{figure*}

To test this, we constructed a flux-flux diagram, shown in the left panel of Fig.~\ref{fig:galaxy_spec}, to decompose the spectrum between a variable (AGN) and a fixed component \citep[host galaxy; e.g.,][]{Fausnaugh:2016,Starkey:2017,Cackett:2020}. We used the driving light curve 
$X(t)$ obtained in Section~\ref{sec:cream} which is normalised so that
$\left<X\right>=0$ and $\left<X^2\right>=1$. Then, we constructed a simple model for each light curve following:
\begin{equation}
F(\lambda) = C(\lambda) + S(\lambda)\, X(t),
\end{equation}
where the flux at each epoch $t$, is a linear combination of $C(\lambda)$ and $S(\lambda)$. The correlation between the components throughout the luminosity range (noted as AGN low and AGN high) is well described by the linear model throughout the flux range of our sample, showing no evidence for curvature. Thus, extrapolating the linear model of the shortest wavelength until it reaches zero flux (to within 1$\sigma$), provides us with a lower limit on the galaxy and DCE contribution. Taking a vertical slice in this limit (dot dashed line in Fig.~\ref{fig:galaxy_spec}), we obtain the spectrum of the non-variable component of the light that we attribute to the underlying host galaxy. The de-reddened SED of the different components (low-state, high-state AGN and Galaxy spectrum) are explicitly shown in the bottom panel of Fig.~\ref{fig:galaxy_spec}. All flux values shown are de-reddened using  E(B-V)$=0.026\pm0.001$ \citep{Schlafly:2011} and the \citet{Fitzpatrick:1999} model for Galactic dust extinction. 
As a sanity check, we have included the infrared measurements \citep{vo:sasmirala_cone,Asmus:2014MNRAS}, shown as open circles in Fig.~\ref{fig:galaxy_spec}. These are consistent with the estimated galactic contribution.
Using $V$ band as a proxy for the 5100~\AA\ monochromatic flux, we find an average disc luminosity of $\log(\lambda  F_{\lambda}/\, {\rm erg\, s}^{-1})=44.1$, slightly higher (0.18 dex) than that obtained from spectro-photometric decomposition $\log(L_{5100} /\, {\rm erg\, s}^{-1})=43.92$ \citep{Bentz:2013}. This difference however is within the variability observed in this campaign and could explain the discrepancy.

The AGN spectrum shows a power-law slightly bluer than the canonical disc SED where $F_{\nu}\propto\nu^{1/3} \propto\lambda^{-1/3}$. 
The observed disc spectral index is $\alpha=0.50\pm0.03$ for the variable component AGN using all bands and a shallower value when omitting U-band $\alpha=0.45\pm0.05$ (shown as the black and grey lines of Fig.~\ref{fig:galaxy_spec}). It has been noted that the variable component in AGN is contaminated by other additional elements such as the BLR \citep{Chelouche:2019} and emission line contribution. Thus, the observed variable SED may not be the pure accretion disc component and thus deviations/bias are expected. We show this explicitly in the right panel of Fig.~\ref{fig:galaxy_spec} where we display the predicted SED for a standard thin disc in \fair. In general, the predicted SED is a factor $\sim3$ brighter than the average AGN spectrum, a characteristic previously observed in other systems e.g., NGC 5548 \citep{Starkey:2017}. 

\subsection{Diffuse Continuum Emission from the BLR}\label{sec:dce}

The spectral decomposition and the lag spectrum both show evidence for deviation from simple power-laws. In particular the \textit{U}-band excess has been previously observed in the \swift\ sample of 5 AGN \citep{Edelson:2019} and more clearly shown in the HST data of NGC~4593 \citep{Cackett:2018} and NGC~5548 \citep{Fausnaugh:2016}. Furthermore, non-disc contributions to the optical reverberation signal of Mrk~279 across all continuum bands are needed to fully account for the variability observed \citep{Chelouche:2019}.
These discrepancies have been interpreted as the contribution from the DCE of the BLR, due to free-free and free-bound hydrogen transitions \citep[e.g.,][]{Korista:2001}. The imprint of the BLR in the lag spectrum is particularly reflected as longer lags towards the Balmer (3646 \AA) and Paschen (8204 \AA) edges in the optical range \citep{Lawther:2018,Korista:2019,Netzer:2020}. For local AGN , such longer lags should be particularly noticeable in $U$ and $i$ bands, respectively (where redshift is not large enough to move the edges out of the filter passbands).

The $U$-band excess is observed in data from both \swift\ and \LCO\ during this first year of \fair\ monitoring.
The excess $U$-band lag relative to the $\tau\propto\lambda^{4/3}$ relation is quite small, $\sim20\%$, which limits the extent to which reprocessing in the BLR can contribute to the observed variations. In addition to the lag spectrum, the variable component of the SED also presents evidence for the Balmer edge and marginal evidence in the Paschen edge in emission. 

We present here a physically-motivated fit to the observed lag spectrum based on the previous work of \citet{Korista:2019}, who modelled the well-studied case of NGC~5548. We use the 
DCE lag spectrum contribution from the BLR (top panel in their figure~6) and scale this as recommended to reflect the luminosity difference between NGC~5548 and \fair, hence the size of the BLR, following the radius-luminosity relation $R_{\rm BLR}\propto L^{1/2}$, where the ratio between their luminosity is ($L_{\rm AGN}/L_{\rm NGC\,5548}$) = (0.02/0.1). 
We then mix the disk lag spectrum
$\tau_{\rm disc}(\lambda)$
and the luminosity-scaled DCE lag spectrum $\tau_{\rm DCE}(\lambda)$,
weighted at each wavelength by their relative fluxes.
For NGC~5548, the wavelength-dependent flux ratio $f(\lambda)$ between the DCE and the disc emission
are given by \citet[][ see their Fig.~9]{Korista:2019}. 
Therefore, the lag spectrum is defined by the following parametrisation:
\begin{equation}
    \tau(\lambda) -\tau(\lambda_0) = \frac{
    \tau_{\rm disc} \left( \lambda/\lambda_0\right)^\alpha + \tau_{\rm DCE}(\lambda) \, B\,  f(\lambda)}{1+ B  f(\lambda)} - \tau_{0}\ ,
\end{equation}
where $\tau_{\rm disk}$ is the disk continuum lag at $\lambda=\lambda_0$, $\alpha$ is the power-law index
of the disk lag spectrum, $\tau_{\rm DCE}(\lambda)$ is the DCE lag spectrum
of NGC~5548, scaled by $L^{1/2}$ to the luminosity of \fair, $f(\lambda)$ is the flux ratio in NGC~5548 of the DCE spectrum to the disk spectrum, $B$ is a dilution factor,
allowing the DCE contribution to be stronger or weaker in \fair\ than in NGC~5548, and
$\tau_0$ is the total (disk+DCE) lag at the reference wavelength, $\lambda_0=1928$ \AA\ for our CCF lags.

\begin{figure}
	\includegraphics[trim={0cm 0.1cm 0cm 0cm}, clip,width=1.\columnwidth]{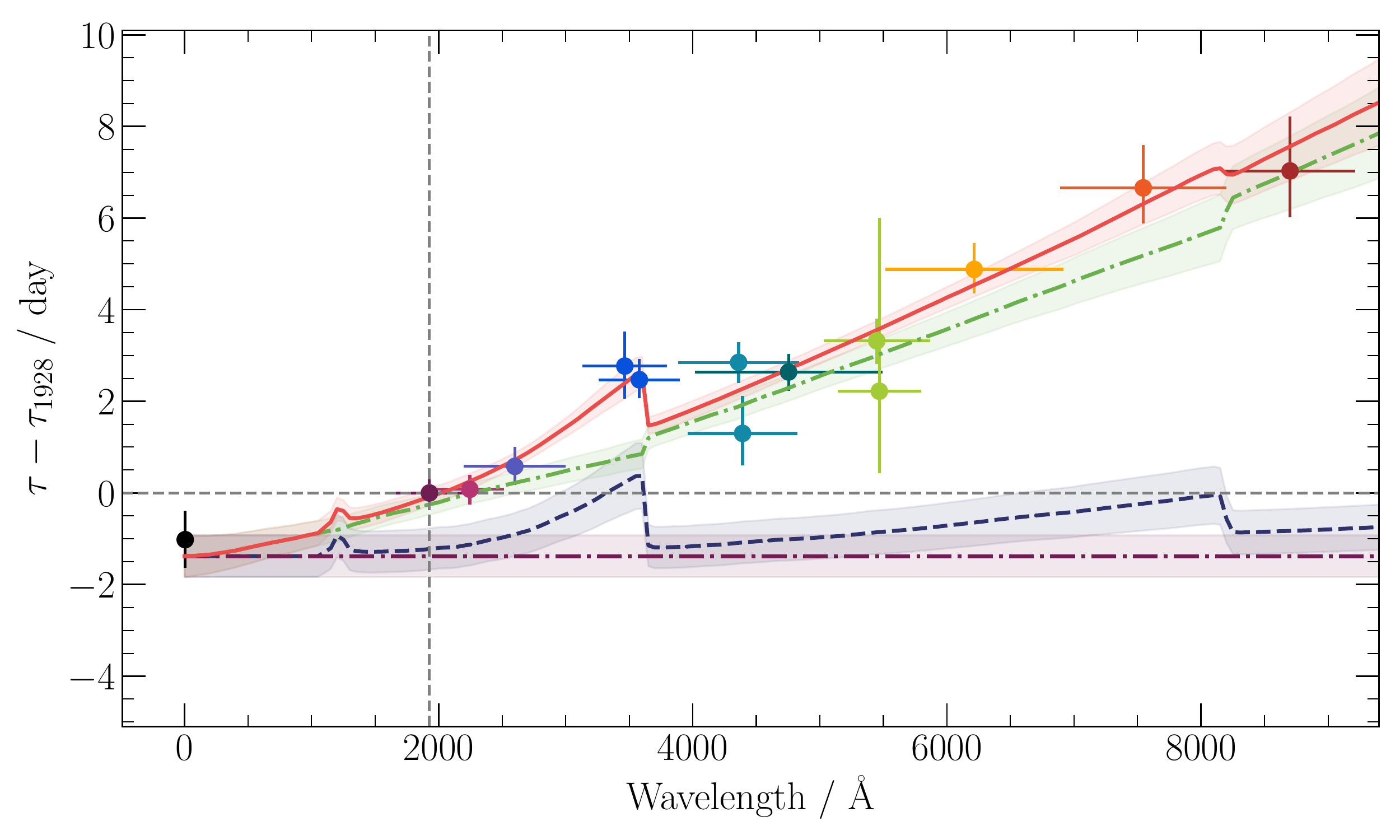}
    \caption{Lag spectrum fit using a scaled version of the DCE model for NGC 5548 (blue dashed line), the accretion disc model represented by a power law (green dashed line). The UVW2 offset is shown as the straight purple line. The envelope in each line represent the 1$\sigma$ confidence intervals.}
    \label{fig:kg19_fit}
\end{figure}

Our 4-parameter model fits for  $\alpha$, $\tau_{\rm disc}$, $B$, and $\tau_0$, with
$\tau_{\rm DCE}(\lambda)$ and $f(\lambda)$ taken from \citet{Korista:2019}. We used a MCMC procedure to sample the posterior parameter distribution using {\sc emcee} \citep{Foreman-Mackey:2013}, using uniform priors for all parameters. 
We find marginalised posterior distributions for each parameters to be $\alpha=1.39\pm0.27$~days, $\tau_{\rm disc}=1.08\pm0.44$~days, $B=0.26\pm0.12$, and $\tau_0=1.31\pm0.48$~days.
The resulting fit, and contributions from each component with their respective 1$\sigma$ error envelope, are shown in Fig.~\ref{fig:kg19_fit}. Joint and marginal posterior  distributions for the parameters are shown in Fig.~\ref{fig:A2}.
 
Notice that for this disc+DCE model of the lag spectrum of \fair, the DCE enhances lags near the Balmer and Paschen edges without greatly influencing the power-law disk lag spectrum.
Thus the simpler procedure of omitting the $U$-band lag when fitting a power-law lag spectrum seems to be justified \citep[as suggested by][]{Korista:2001}. If this is true also for AGN that have measured disc continuum lags larger than expected, then the DCE component may not cure the problem. 

The previous analysis shows that, qualitatively, the inclusion of the DCE provides a better representation of the measured lag spectrum. 
In particular, our parametrisation uses a simple scaling of AGN NGC 5548 as a proxy for the DCE contribution. This assumes that the wavelength dependence remains the same for \fair\ while and the main difference lies on the fractional contribution (i.e., scaling). However, changes in the ionising luminosity between these two systems can give rise to significant changes in both of aspects. For example, specific differences in the ionising spectral shape can provide changes in the DCE contribution by a factor $\sim2$ \citep{Netzer:2020}.
A detailed modelling of the DCE in \fair\ and their effect on the lag spectrum and SED is beyond the scope of this paper and will be deferred to a future analysis including additional spectroscopic and photometric monitoring over the 2-year Key Project.
 
\subsection{Fast and Slow Variations}\label{sec:slow}

\begin{figure*}
	\includegraphics[width=17cm]{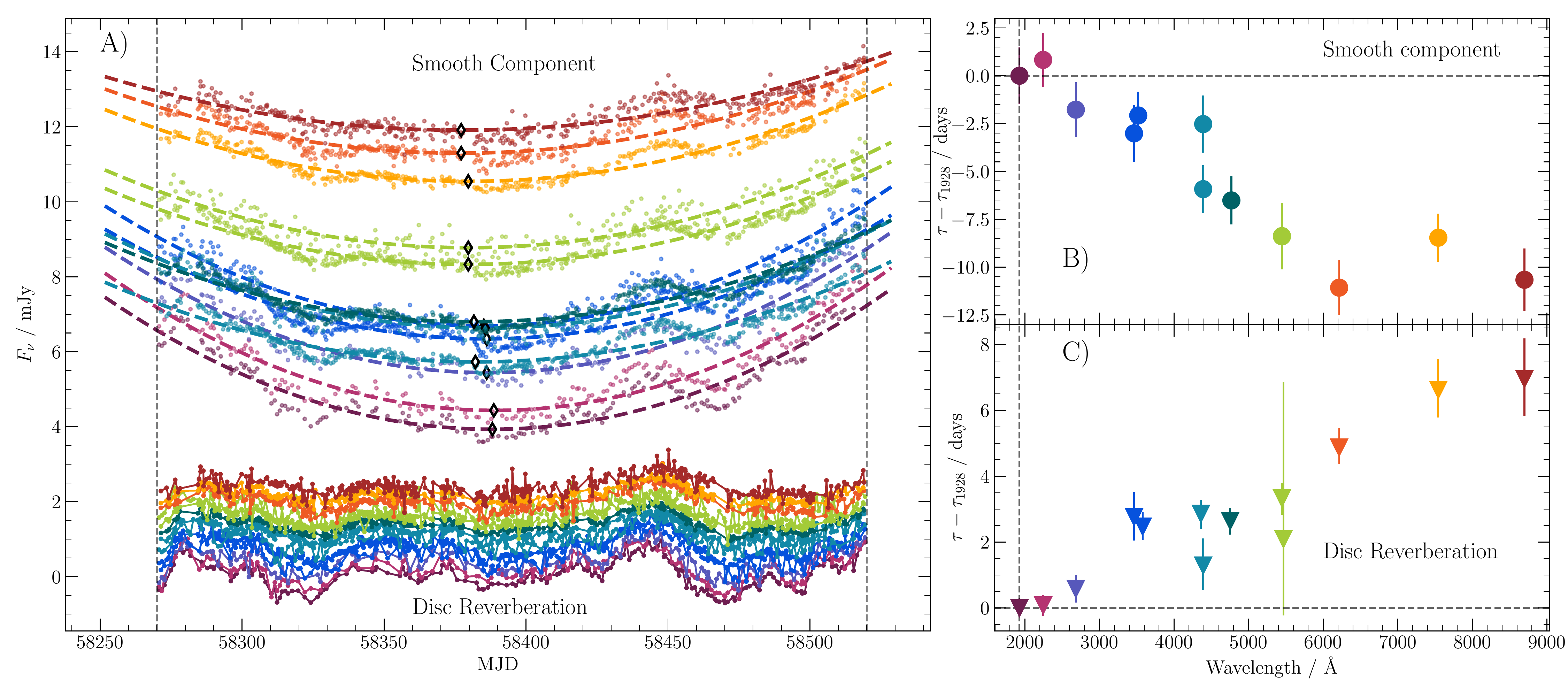}
    \caption{Identification of two variable components in the Year~1 data of \fair. {\bf A)} 
    shows the \LCO\ and \swift\ light curves, colour coded by wavelength, with their respective parabolic fit as the dashed line. The minimum of each parabola is shown as a black diamond. Note that these progress to earlier times in the redder light curves. Residual light curves (data minus parabolic fit) are shown in the bottom of the panel, with an arbitrary shift for display purposes. 
    Note the peak near MJD 58450 shifting to later times in the redder light curves.
    {\bf B)} The blue lag spectrum of the slow variations, as measured by the minimum of the quadratic fit used to detrend each individual light curve. We show the difference between these minima relative to the UVW2-band.
    {\bf C)} The red lag spectrum of the fast variations, as measured from the CCF analysis (see Section~\ref{sec:ccf}) of the residual light curves.
    These lags increasing with wavelength are attributed to light-travel times for X-rays reprocessed in small hot and later in large cool annuli of the accretion disc.
    }
    \label{fig:quad_fit_lag}
\end{figure*}
Long-term optical correlations have been observed in several AGN \citep[e.g.,][]{Uttley:2003,Breedt:2009}. These studies concluded that the X-ray reprocessing was not the sole driver of variability. 
The decomposition of the driving light curve from {\sc cream} (see Fig.~\ref{fig:driving_lc}) suggests that the observed variability arises from rapid variations, likely due to reprocessing of X-rays, and slower variations on longer timescales ($>100$ days).
To quantify this slow component, we use a quadratic fit to detrend the light curves, as performed in Section \ref{sec:ccf}. Since this fit may carry temporal information, we have parametrised it as follows to characterise its evolution:
\begin{equation}
    F_{\nu}(\lambda,t) = F_0(\lambda) + \Delta F(\lambda) \, \left(\frac{t-t_0(\lambda)}{100\, {\rm days}}\right)^2,
\end{equation}
where $F_0$ is the flux at the minimum of the
parabola, which occurs at time $t_0$,
and $\Delta F$ is the flux rise 100 days 
before or after the minimum.
In the left panel of Fig.~\ref{fig:quad_fit_lag} we show the full decomposition of \swift\ and \lco\ light curves. The fast variations (residuals after detrending)
have a red lag spectrum.
The lag increases with wavelength,
as obtained from the CCF lags (see Section~\ref{sec:ccf}) and we reproduce in the upper right panel of Fig.~\ref{fig:quad_fit_lag}.
For the parabolic fits, the times of minimum, $t_0(\lambda)$ 
(shown as a diamond on each light curve in Fig.~\ref{fig:quad_fit_lag}) 
decrease with wavelength. Using the UVW2 filter as reference, we show this distribution of these minima in the lower right panel of Fig.~\ref{fig:quad_fit_lag}. While the uncertainties are larger than the reverberation signal, the trend is clear for most wavelengths despite the flattening at the two longest ones. We explore possible origins of this component in the following section.

This opposite lag signal in the slow component could be interpreted as accretion-rate fluctuations propagating inwards through the accretion disc,
from cooler to hotter regions,
on the local viscous timescale \citep[e.g.,][]{Lyubarskii:1997,Arevalo:2008}. This would translate to an observable wavelength-dependent lag where the lower energies lead the higher ones, such as the trend shown in Fig.~\ref{fig:quad_fit_lag}.
However, the $\sim10$-day timescale indicated by the time shifts in the minima of the parabolic fits is rather shorter than the viscous timescale expected for classical thin discs. However, this delay is not simply the propagation time of accretion fluctuations, but propagation time convolved with the weighted radial emission profile in each band. For any given radius the disc will emit over a range of wavelengths which will further suppress the lag, since one is unable to map one particular filter/band to a narrow range of radii.

Moreover, the variable emission observed at any given band may carry information at different time-scales, which trace different average radii on the disc \citep[e.g.,][]{Arevalo:2008}. For example, the inner radii of the disc may only produce a small fraction of optical light (due to smaller centrally concentrated regions closer to the BH) and will mostly vary (slowly) due to accretion fluctuations propagating at a viscous timescale. While the outer radii may produce a larger fraction of optical light which is relatively constant (since the timescales are longer farther out), a fraction of its light will vary due to X-ray reprocessing.
Thus the fast variability may come from larger radii than the slow variability (hence the lags can be similar but with different sign).
In BH X-ray binaries, there is strong evidence that thermal emission from the innermost parts of the optically thick disc also shows strong intrinsic variability (at lower energies) that leads the Comptonised (higher-energy) emission by time-delays much longer than light-travel times \citep{Uttley:2011}. In particular, they observe lags of $\sim0.1$ s  on seconds time-scales, which scaled for \fair\ would represent few tens of days lead on years time-scales.

An alternative way to interpret the parabolic fits is that a change in colour of the slow component's SED can mimic time-delayed variations. If the SED is slightly redder at the start of our observations and bluer near the end, the minimum of the parabola will be later in the blue and earlier in the red,  mimicking a time delay increasing from red to blue. 
In this interpretation, the relevant timescale is not the 10-day shift of the minimum of the parabola, but the much longer time over which the SED colour changes. 

Multi-year monitoring of \fair\ should clarify which if either of these interpretations is correct and show the robustness the detection of the slow component.
The simple parabolic fit, while informative in this particular year of monitoring, is obviously not viable as a long-term description of the slow component. This issue will require to fully model the underlying background contribution as a smooth polynomial or as an independent light curve (with different PSD as the reverberation signal). In addition, if these components have indeed opposite lag directions, the ``average" lag measured without any detrending (that for \fair\ removes the slow component) might produce bias lag measurements. This could potentially affect all reverberation experiments if not considered. Our continued monitoring of \fair\ will provide a more detailed record of the slow and fast variations, allowing for a more secure separation and characterisation of the fast and slow components over a longer time span and allow us to understand the correlations between them.


\section{Conclusions}\label{sec:conclusions}

We report analysis of the first year of X-ray, UV and optical monitoring of \fair\ with the \swift\ Observatory and the ground-based \lco\ robotic telescope network.
Sustained photometric monitoring of the X-rays and 13 UV/optical bands
was achieved with a 1-day cadence by \swift\ and sub-day cadence by \LCO.

The main results of our analysis of the spectral variations may be summarised as follows:

1. The observed UV and optical light curves can be decomposed into slow variations that we model with a parabola, and faster variations that correlate with the observed X-ray variations.

2. Cross-correlation lag measurements for the faster variations, measured relative to those in the \swift\ UVW2 band, increase with wavelength with a power-law index $\alpha=1.26\pm0.23$, compatible with the $\tau\propto\lambda^{4/3}$ prediction of accretion disc theory.

3. The lag spectrum relative to UVW2 increases from $-1.0\pm0.6$ for the X-rays to $7.0\pm1.2$ days at longest wavelength \emph{$z_s$} band, compatible with the prediction of accretion disc theory for the source Eddington ratio $L/L_{\rm edd}=0.02$.

4. Decomposing the data into mean and variable components, we isolate the disc spectrum as the variable component and find that its power-law spectral index $0.5\pm0.02$,
is slightly bluer than the standard accretion disc prediction of 
$F_\nu\propto\nu^{1/3}$.

5. Evidence for a Balmer jump in emission is evident in both the delay spectrum and the
disc flux spectrum, with a small $\sim20\%$ amplitude relative to the power-law. This places limits on the contribution of diffuse bound-free continuum
emission from reprocessing in the broad-line region.

6. The slowly varying component shows a blue lag spectrum, with lags decreasing with wavelength. These may be indicative of variations in the accretion flow in the disc. However, the origin of the slow component is uncertain but should be clarified through our continued monitoring of \fair.

\section*{Acknowledgements}
We would like to thank the anonymous referee for their comments that greatly improved the manuscript.
JVHS and KH acknowledge support from STFC grant ST/R000824/1.
RE gratefully acknowledges support from NASA \swift\ Key Project grant number 80NSSC19K0153.
J.M.G. gratefully acknowledges support from NASA under the ADAP award 80NSSC17K0126.
AAB, KLP and PAE acknowledge support from the UK Space Agency.
Research by AJB was supported by NSF grant AST-1907290.
EMC gratefully acknowledges support from the NSF through grant AST-1909199.
MV gratefully acknowledges financial support from the Independent Research Fund Denmark via grant number DFF  8021-00130.
The authors appreciate the hard work and dedication of the \swift\ Observatory staff, who created a new UVOT mode in support this project and put in considerable effort in scheduling this large program.
This work makes use of observations from the \lco\ network, and of the NASA/ IPAC Infrared Science Archive, which is operated by the Jet Propulsion Laboratory, California Institute of Technology, under contract with the National Aeronautics and Space Administration. 
This research was made possible through the use of the AAVSO Photometric All-Sky Survey (APASS), funded by the Robert Martin Ayers Sciences Fund and NSF AST-1412587.
This research also made use of {\sc astropy}, a community-developed core {\sc python} package for Astronomy \citep{Astropy-Collaboration:2013aa} and {\sc matplotlib} \citep{Hunter:2007aa}.

\section*{Data availability}
The raw datasets were derived from sources in the public domain: LCO archive \url{https://archive.lco.global} and {\it Swift} archive \url{https://www.swift.ac.uk/swift_live}. The inter-calibrated light curves and Swift detector maps are available in Zenodo, at~\url{https://dx.doi.org/10.5281/zenodo.3956577}.


\bibliographystyle{mnras}
\bibliography{references} 

\begin{thebibliography}{}
\makeatletter
\relax
\def\mn@urlcharsother{\let\do\@makeother \do\$\do\&\do\#\do\^\do\_\do\%\do\~}
\def\mn@doi{\begingroup\mn@urlcharsother \@ifnextchar [ {\mn@doi@}
  {\mn@doi@[]}}
\def\mn@doi@[#1]#2{\def\@tempa{#1}\ifx\@tempa\@empty \href
  {http://dx.doi.org/#2} {doi:#2}\else \href {http://dx.doi.org/#2} {#1}\fi
  \endgroup}
\def\mn@eprint#1#2{\mn@eprint@#1:#2::\@nil}
\def\mn@eprint@arXiv#1{\href {http://arxiv.org/abs/#1} {{\tt arXiv:#1}}}
\def\mn@eprint@dblp#1{\href {http://dblp.uni-trier.de/rec/bibtex/#1.xml}
  {dblp:#1}}
\def\mn@eprint@#1:#2:#3:#4\@nil{\def\@tempa {#1}\def\@tempb {#2}\def\@tempc
  {#3}\ifx \@tempc \@empty \let \@tempc \@tempb \let \@tempb \@tempa \fi \ifx
  \@tempb \@empty \def\@tempb {arXiv}\fi \@ifundefined
  {mn@eprint@\@tempb}{\@tempb:\@tempc}{\expandafter \expandafter \csname
  mn@eprint@\@tempb\endcsname \expandafter{\@tempc}}}

\bibitem[\protect\citeauthoryear{{Ar{\'e}valo}, {Uttley}, {Kaspi}, {Breedt},
  {Lira}  \& {McHardy}}{{Ar{\'e}valo} et~al.}{2008}]{Arevalo:2008}
{Ar{\'e}valo} P.,  {Uttley} P.,  {Kaspi} S.,  {Breedt} E.,  {Lira} P.,
  {McHardy} I.~M.,  2008, \mn@doi [\mnras] {10.1111/j.1365-2966.2008.13719.x},
  \href {https://ui.adsabs.harvard.edu/abs/2008MNRAS.389.1479A} {389, 1479}

\bibitem[\protect\citeauthoryear{Asmus, Hoenig, Gandhi, Smette  \&
  Duschl}{Asmus et~al.}{2013}]{vo:sasmirala_cone}
Asmus D.,  Hoenig S.,  Gandhi P.,  Smette A.,   Duschl W.,  2013, Subarcsecond
  mid-infrared atlas of local {AGN}, Cone Search, {VO} resource provided by the
  {GAVO} Data Center, \url
  {http://dc.zah.uni-heidelberg.de/sasmirala/q/cone/info}

\bibitem[\protect\citeauthoryear{{Asmus}, {H{\"o}nig}, {Gandhi}, {Smette}  \&
  {Duschl}}{{Asmus} et~al.}{2014}]{Asmus:2014MNRAS}
{Asmus} D.,  {H{\"o}nig} S.~F.,  {Gandhi} P.,  {Smette} A.,   {Duschl} W.~J.,
  2014, \mn@doi [\mnras] {10.1093/mnras/stu041}, \href
  {https://ui.adsabs.harvard.edu/abs/2014MNRAS.439.1648A} {439, 1648}

\bibitem[\protect\citeauthoryear{{Astropy Collaboration} et~al.,}{{Astropy
  Collaboration} et~al.}{2013}]{Astropy-Collaboration:2013aa}
{Astropy Collaboration} et~al., 2013, \aap, 558, A33

\bibitem[\protect\citeauthoryear{{Bentz} \& {Katz}}{{Bentz} \&
  {Katz}}{2015}]{Bentz:2015}
{Bentz} M.~C.,  {Katz} S.,  2015, \mn@doi [\pasp] {10.1086/679601}, \href
  {https://ui.adsabs.harvard.edu/abs/2015PASP..127...67B} {127, 67}

\bibitem[\protect\citeauthoryear{{Bentz} et~al.,}{{Bentz}
  et~al.}{2013}]{Bentz:2013}
{Bentz} M.~C.,  et~al., 2013, \mn@doi [\apj] {10.1088/0004-637X/767/2/149},
  \href {https://ui.adsabs.harvard.edu/abs/2013ApJ...767..149B} {767, 149}

\bibitem[\protect\citeauthoryear{{Bertin} \& {Arnouts}}{{Bertin} \&
  {Arnouts}}{1996}]{bertin:1996}
{Bertin} E.,  {Arnouts} S.,  1996, \mn@doi [\aaps] {10.1051/aas:1996164}, \href
  {https://ui.adsabs.harvard.edu/abs/1996A&AS..117..393B} {117, 393}

\bibitem[\protect\citeauthoryear{{Blackburn}}{{Blackburn}}{1995}]{Blackburn:1995}
{Blackburn} J.~K.,  1995, {FTOOLS: A FITS Data Processing and Analysis Software
  Package}.
p.~367

\bibitem[\protect\citeauthoryear{{Blandford} \& {McKee}}{{Blandford} \&
  {McKee}}{1982}]{Blandford:1982}
{Blandford} R.~D.,  {McKee} C.~F.,  1982, \mn@doi [\apj] {10.1086/159843},
  \href {https://ui.adsabs.harvard.edu/abs/1982ApJ...255..419B} {255, 419}

\bibitem[\protect\citeauthoryear{{Brandt} \& {Alexander}}{{Brandt} \&
  {Alexander}}{2015}]{Brandt:2015}
{Brandt} W.~N.,  {Alexander} D.~M.,  2015, \mn@doi [\aapr]
  {10.1007/s00159-014-0081-z}, \href
  {https://ui.adsabs.harvard.edu/abs/2015A&ARv..23....1B} {23, 1}

\bibitem[\protect\citeauthoryear{{Breedt} et~al.,}{{Breedt}
  et~al.}{2009}]{Breedt:2009}
{Breedt} E.,  et~al., 2009, \mn@doi [\mnras]
  {10.1111/j.1365-2966.2008.14302.x}, \href
  {https://ui.adsabs.harvard.edu/abs/2009MNRAS.394..427B} {394, 427}

\bibitem[\protect\citeauthoryear{{Breeveld}, {Landsman}, {Holland}, {Roming},
  {Kuin}  \& {Page}}{{Breeveld} et~al.}{2011}]{Breeveld:2011}
{Breeveld} A.~A.,  {Landsman} W.,  {Holland} S.~T.,  {Roming} P.,  {Kuin}
  N.~P.~M.,   {Page} M.~J.,  2011, in {McEnery} J.~E.,  {Racusin} J.~L.,
  {Gehrels} N.,  eds,  American Institute of Physics Conference Series Vol.
  1358, American Institute of Physics Conference Series. pp 373--376
  (\mn@eprint {arXiv} {1102.4717}), \mn@doi{10.1063/1.3621807}

\bibitem[\protect\citeauthoryear{{Brown} et~al.,}{{Brown}
  et~al.}{2013}]{Brown:2013}
{Brown} T.~M.,  et~al., 2013, \mn@doi [\pasp] {10.1086/673168}, \href
  {https://ui.adsabs.harvard.edu/abs/2013PASP..125.1031B} {125, 1031}

\bibitem[\protect\citeauthoryear{{Burrows} et~al.,}{{Burrows}
  et~al.}{2005}]{Burrows:2005}
{Burrows} D.~N.,  et~al., 2005, \mn@doi [\ssr] {10.1007/s11214-005-5097-2},
  \href {https://ui.adsabs.harvard.edu/abs/2005SSRv..120..165B} {120, 165}

\bibitem[\protect\citeauthoryear{{Cackett}, {Horne}  \& {Winkler}}{{Cackett}
  et~al.}{2007}]{Cackett:2007}
{Cackett} E.~M.,  {Horne} K.,   {Winkler} H.,  2007, \mn@doi [\mnras]
  {10.1111/j.1365-2966.2007.12098.x}, \href
  {https://ui.adsabs.harvard.edu/abs/2007MNRAS.380..669C} {380, 669}

\bibitem[\protect\citeauthoryear{{Cackett}, {Chiang}, {McHardy}, {Edelson},
  {Goad}, {Horne}  \& {Korista}}{{Cackett} et~al.}{2018}]{Cackett:2018}
{Cackett} E.~M.,  {Chiang} C.-Y.,  {McHardy} I.,  {Edelson} R.,  {Goad} M.~R.,
  {Horne} K.,   {Korista} K.~T.,  2018, \mn@doi [\apj]
  {10.3847/1538-4357/aab4f7}, \href
  {https://ui.adsabs.harvard.edu/abs/2018ApJ...857...53C} {857, 53}

\bibitem[\protect\citeauthoryear{{Cackett} et~al.,}{{Cackett}
  et~al.}{2020}]{Cackett:2020}
{Cackett} E.~M.,  et~al., 2020, arXiv e-prints, \href
  {https://ui.adsabs.harvard.edu/abs/2020arXiv200503685C} {p. arXiv:2005.03685}

\bibitem[\protect\citeauthoryear{{Chelouche}, {Pozo Nu{\~n}ez}  \&
  {Kaspi}}{{Chelouche} et~al.}{2019}]{Chelouche:2019}
{Chelouche} D.,  {Pozo Nu{\~n}ez} F.,   {Kaspi} S.,  2019, \mn@doi [Nature
  Astronomy] {10.1038/s41550-018-0659-x}, \href
  {https://ui.adsabs.harvard.edu/abs/2019NatAs...3..251C} {3, 251}

\bibitem[\protect\citeauthoryear{{De Rosa} et~al.,}{{De Rosa}
  et~al.}{2015}]{deRosa:2015}
{De Rosa} G.,  et~al., 2015, \mn@doi [\apj] {10.1088/0004-637X/806/1/128},
  \href {https://ui.adsabs.harvard.edu/abs/2015ApJ...806..128D} {806, 128}

\bibitem[\protect\citeauthoryear{{Degenaar}, {Miller}, {Kennea}, {Gehrels},
  {Reynolds}  \& {Wijnands}}{{Degenaar} et~al.}{2013}]{Degenaar:2013}
{Degenaar} N.,  {Miller} J.~M.,  {Kennea} J.,  {Gehrels} N.,  {Reynolds} M.~T.,
    {Wijnands} R.,  2013, \mn@doi [\apj] {10.1088/0004-637X/769/2/155}, \href
  {https://ui.adsabs.harvard.edu/abs/2013ApJ...769..155D} {769, 155}

\bibitem[\protect\citeauthoryear{{Edelson} et~al.,}{{Edelson}
  et~al.}{2015}]{Edelson:2015}
{Edelson} R.,  et~al., 2015, \mn@doi [\apj] {10.1088/0004-637X/806/1/129},
  \href {https://ui.adsabs.harvard.edu/abs/2015ApJ...806..129E} {806, 129}

\bibitem[\protect\citeauthoryear{{Edelson} et~al.,}{{Edelson}
  et~al.}{2017}]{Edelson:2017}
{Edelson} R.,  et~al., 2017, \mn@doi [\apj] {10.3847/1538-4357/aa6890}, \href
  {https://ui.adsabs.harvard.edu/abs/2017ApJ...840...41E} {840, 41}

\bibitem[\protect\citeauthoryear{{Edelson} et~al.,}{{Edelson}
  et~al.}{2019}]{Edelson:2019}
{Edelson} R.,  et~al., 2019, \mn@doi [\apj] {10.3847/1538-4357/aaf3b4}, \href
  {https://ui.adsabs.harvard.edu/abs/2019ApJ...870..123E} {870, 123}

\bibitem[\protect\citeauthoryear{{Emmanoulopoulos}, {Papadakis}, {McHardy},
  {Nicastro}, {Bianchi}  \& {Ar{\'e}valo}}{{Emmanoulopoulos}
  et~al.}{2011}]{Emmanoulopoulos:2011}
{Emmanoulopoulos} D.,  {Papadakis} I.~E.,  {McHardy} I.~M.,  {Nicastro} F.,
  {Bianchi} S.,   {Ar{\'e}valo} P.,  2011, \mn@doi [\mnras]
  {10.1111/j.1365-2966.2011.18834.x}, \href
  {https://ui.adsabs.harvard.edu/abs/2011MNRAS.415.1895E} {415, 1895}

\bibitem[\protect\citeauthoryear{{Evans} et~al.,}{{Evans}
  et~al.}{2009}]{Evans:2009}
{Evans} P.~A.,  et~al., 2009, \mn@doi [\mnras]
  {10.1111/j.1365-2966.2009.14913.x}, \href
  {https://ui.adsabs.harvard.edu/abs/2009MNRAS.397.1177E} {397, 1177}

\bibitem[\protect\citeauthoryear{{Event Horizon Telescope Collaboration}
  et~al.,}{{Event Horizon Telescope Collaboration}
  et~al.}{2019}]{EventHorizon:2019}
{Event Horizon Telescope Collaboration} et~al., 2019, \mn@doi [\apjl]
  {10.3847/2041-8213/ab0ec7}, \href
  {https://ui.adsabs.harvard.edu/abs/2019ApJ...875L...1E} {875, L1}

\bibitem[\protect\citeauthoryear{{Fabian}}{{Fabian}}{2012}]{Fabian:2012}
{Fabian} A.~C.,  2012, \mn@doi [\araa] {10.1146/annurev-astro-081811-125521},
  \href {https://ui.adsabs.harvard.edu/abs/2012ARA&A..50..455F} {50, 455}

\bibitem[\protect\citeauthoryear{{Fairall}}{{Fairall}}{1977}]{Fairall:1977}
{Fairall} A.~P.,  1977, \mn@doi [\mnras] {10.1093/mnras/180.3.391}, \href
  {https://ui.adsabs.harvard.edu/abs/1977MNRAS.180..391F} {180, 391}

\bibitem[\protect\citeauthoryear{{Fausnaugh} et~al.,}{{Fausnaugh}
  et~al.}{2016}]{Fausnaugh:2016}
{Fausnaugh} M.~M.,  et~al., 2016, \mn@doi [\apj] {10.3847/0004-637X/821/1/56},
  \href {https://ui.adsabs.harvard.edu/abs/2016ApJ...821...56F} {821, 56}

\bibitem[\protect\citeauthoryear{{Fitzpatrick}}{{Fitzpatrick}}{1999}]{Fitzpatrick:1999}
{Fitzpatrick} E.~L.,  1999, \mn@doi [\pasp] {10.1086/316293}, \href
  {https://ui.adsabs.harvard.edu/abs/1999PASP..111...63F} {111, 63}

\bibitem[\protect\citeauthoryear{Foreman-Mackey}{Foreman-Mackey}{2016}]{corner}
Foreman-Mackey D.,  2016, \mn@doi [The Journal of Open Source Software]
  {10.21105/joss.00024}, 24

\bibitem[\protect\citeauthoryear{{Foreman-Mackey}, {Hogg}, {Lang}  \&
  {Goodman}}{{Foreman-Mackey} et~al.}{2013}]{Foreman-Mackey:2013}
{Foreman-Mackey} D.,  {Hogg} D.~W.,  {Lang} D.,   {Goodman} J.,  2013, \mn@doi
  [\pasp] {10.1086/670067}, \href
  {https://ui.adsabs.harvard.edu/abs/2013PASP..125..306F} {125, 306}

\bibitem[\protect\citeauthoryear{{Gardner} \& {Done}}{{Gardner} \&
  {Done}}{2017}]{Gardner:2017}
{Gardner} E.,  {Done} C.,  2017, \mn@doi [\mnras] {10.1093/mnras/stx946}, \href
  {https://ui.adsabs.harvard.edu/abs/2017MNRAS.470.3591G} {470, 3591}

\bibitem[\protect\citeauthoryear{{Gaskell} \& {Peterson}}{{Gaskell} \&
  {Peterson}}{1987}]{Gaskell:1987}
{Gaskell} C.~M.,  {Peterson} B.~M.,  1987, \mn@doi [\apjs] {10.1086/191216},
  \href {https://ui.adsabs.harvard.edu/abs/1987ApJS...65....1G} {65, 1}

\bibitem[\protect\citeauthoryear{{Gehrels} et~al.,}{{Gehrels}
  et~al.}{2004}]{Gehrels:2004}
{Gehrels} N.,  et~al., 2004, \mn@doi [\apj] {10.1086/422091}, \href
  {https://ui.adsabs.harvard.edu/abs/2004ApJ...611.1005G} {611, 1005}

\bibitem[\protect\citeauthoryear{{Gravity Collaboration} et~al.,}{{Gravity
  Collaboration} et~al.}{2018}]{Gravity:2018}
{Gravity Collaboration} et~al., 2018, \mn@doi [\nat]
  {10.1038/s41586-018-0731-9}, \href
  {https://ui.adsabs.harvard.edu/abs/2018Natur.563..657G} {563, 657}

\bibitem[\protect\citeauthoryear{{Grier} et~al.,}{{Grier}
  et~al.}{2017}]{Grier:2017}
{Grier} C.~J.,  et~al., 2017, \mn@doi [\apj] {10.3847/1538-4357/aa98dc}, \href
  {https://ui.adsabs.harvard.edu/abs/2017ApJ...851...21G} {851, 21}

\bibitem[\protect\citeauthoryear{{Hawley} \& {Phillips}}{{Hawley} \&
  {Phillips}}{1978}]{Hawley:1978}
{Hawley} S.~A.,  {Phillips} M.~M.,  1978, \mn@doi [\apj] {10.1086/156542},
  \href {https://ui.adsabs.harvard.edu/abs/1978ApJ...225..780H} {225, 780}

\bibitem[\protect\citeauthoryear{{Henden}, {Levine}, {Terrell}, {Welch},
  {Munari}  \& {Kloppenborg}}{{Henden} et~al.}{2018}]{henden:2018}
{Henden} A.~A.,  {Levine} S.,  {Terrell} D.,  {Welch} D.~L.,  {Munari} U.,
  {Kloppenborg} B.~K.,  2018, in American Astronomical Society Meeting
  Abstracts \#232. p. 223.06

\bibitem[\protect\citeauthoryear{{Hunter, J. D.}}{{Hunter, J.
  D.}}{2007}]{Hunter:2007aa}
{Hunter, J. D.} 2007, {Computing In Science \& Engineering}, 9, 90

\bibitem[\protect\citeauthoryear{{Korista} \& {Goad}}{{Korista} \&
  {Goad}}{2001}]{Korista:2001}
{Korista} K.~T.,  {Goad} M.~R.,  2001, \mn@doi [\apj] {10.1086/320964}, \href
  {https://ui.adsabs.harvard.edu/abs/2001ApJ...553..695K} {553, 695}

\bibitem[\protect\citeauthoryear{{Korista} \& {Goad}}{{Korista} \&
  {Goad}}{2019}]{Korista:2019}
{Korista} K.~T.,  {Goad} M.~R.,  2019, \mn@doi [\mnras]
  {10.1093/mnras/stz2330}, \href
  {https://ui.adsabs.harvard.edu/abs/2019MNRAS.489.5284K} {489, 5284}

\bibitem[\protect\citeauthoryear{{Lawther}, {Goad}, {Korista}, {Ulrich}  \&
  {Vestergaard}}{{Lawther} et~al.}{2018}]{Lawther:2018}
{Lawther} D.,  {Goad} M.~R.,  {Korista} K.~T.,  {Ulrich} O.,   {Vestergaard}
  M.,  2018, \mn@doi [\mnras] {10.1093/mnras/sty2242}, \href
  {https://ui.adsabs.harvard.edu/abs/2018MNRAS.481..533L} {481, 533}

\bibitem[\protect\citeauthoryear{{Lohfink}, {Reynolds}, {Vasudevan},
  {Mushotzky}  \& {Miller}}{{Lohfink} et~al.}{2014}]{Lohfink:2014}
{Lohfink} A.~M.,  {Reynolds} C.~S.,  {Vasudevan} R.,  {Mushotzky} R.~F.,
  {Miller} N.~A.,  2014, \mn@doi [\apj] {10.1088/0004-637X/788/1/10}, \href
  {https://ui.adsabs.harvard.edu/abs/2014ApJ...788...10L} {788, 10}

\bibitem[\protect\citeauthoryear{{Lohfink} et~al.,}{{Lohfink}
  et~al.}{2016}]{lohfnik:2016}
{Lohfink} A.~M.,  et~al., 2016, \mn@doi [\apj] {10.3847/0004-637X/821/1/11},
  \href {https://ui.adsabs.harvard.edu/abs/2016ApJ...821...11L} {821, 11}

\bibitem[\protect\citeauthoryear{{Lynden-Bell}}{{Lynden-Bell}}{1969}]{Lynden-Bell:1969}
{Lynden-Bell} D.,  1969, \mn@doi [\nat] {10.1038/223690a0}, \href
  {https://ui.adsabs.harvard.edu/abs/1969Natur.223..690L} {223, 690}

\bibitem[\protect\citeauthoryear{{Lyubarskii}}{{Lyubarskii}}{1997}]{Lyubarskii:1997}
{Lyubarskii} Y.~E.,  1997, \mn@doi [\mnras] {10.1093/mnras/292.3.679}, \href
  {https://ui.adsabs.harvard.edu/abs/1997MNRAS.292..679L} {292, 679}

\bibitem[\protect\citeauthoryear{McCully et~al.,}{McCully
  et~al.}{2018}]{curtis_mccully_2018_1257560}
McCully C.,  et~al., 2018, LCOGT/banzai: Initial Release,
  \mn@doi{10.5281/zenodo.1257560}, \url
  {https://doi.org/10.5281/zenodo.1257560}

\bibitem[\protect\citeauthoryear{{McHardy} et~al.,}{{McHardy}
  et~al.}{2018}]{McHardy:2018}
{McHardy} I.~M.,  et~al., 2018, \mn@doi [\mnras] {10.1093/mnras/sty1983}, \href
  {https://ui.adsabs.harvard.edu/abs/2018MNRAS.480.2881M} {480, 2881}

\bibitem[\protect\citeauthoryear{{Netzer}}{{Netzer}}{2020}]{Netzer:2020}
{Netzer} H.,  2020, \mn@doi [\mnras] {10.1093/mnras/staa767}, \href
  {https://ui.adsabs.harvard.edu/abs/2020MNRAS.tmp..704N} {}

\bibitem[\protect\citeauthoryear{{Pal}, {Dewangan}, {Connolly}  \&
  {Misra}}{{Pal} et~al.}{2017}]{Pal:2017}
{Pal} M.,  {Dewangan} G.~C.,  {Connolly} S.~D.,   {Misra} R.,  2017, \mn@doi
  [\mnras] {10.1093/mnras/stw3173}, \href
  {https://ui.adsabs.harvard.edu/abs/2017MNRAS.466.1777P} {466, 1777}

\bibitem[\protect\citeauthoryear{{Peterson}, {Wanders}, {Horne}, {Collier},
  {Alexander}, {Kaspi}  \& {Maoz}}{{Peterson} et~al.}{1998}]{Peterson:1998}
{Peterson} B.~M.,  {Wanders} I.,  {Horne} K.,  {Collier} S.,  {Alexander} T.,
  {Kaspi} S.,   {Maoz} D.,  1998, \mn@doi [\pasp] {10.1086/316177}, \href
  {https://ui.adsabs.harvard.edu/abs/1998PASP..110..660P} {110, 660}

\bibitem[\protect\citeauthoryear{{Peterson} et~al.,}{{Peterson}
  et~al.}{2004}]{Peterson:2004}
{Peterson} B.~M.,  et~al., 2004, \mn@doi [\apj] {10.1086/423269}, \href
  {https://ui.adsabs.harvard.edu/abs/2004ApJ...613..682P} {613, 682}

\bibitem[\protect\citeauthoryear{{Poole} et~al.,}{{Poole}
  et~al.}{2008}]{Poole:2008}
{Poole} T.~S.,  et~al., 2008, \mn@doi [\mnras]
  {10.1111/j.1365-2966.2007.12563.x}, \href
  {https://ui.adsabs.harvard.edu/abs/2008MNRAS.383..627P} {383, 627}

\bibitem[\protect\citeauthoryear{{Pozo Nu{\~n}ez} et~al.,}{{Pozo Nu{\~n}ez}
  et~al.}{2019}]{Pozo-Nunez:2019}
{Pozo Nu{\~n}ez} F.,  et~al., 2019, \mn@doi [\mnras] {10.1093/mnras/stz2830},
  \href {https://ui.adsabs.harvard.edu/abs/2019MNRAS.490.3936P} {490, 3936}

\bibitem[\protect\citeauthoryear{{Ricker}}{{Ricker}}{1978}]{Ricker:1978}
{Ricker} G.~R.,  1978, \mn@doi [\nat] {10.1038/271334a0}, \href
  {https://ui.adsabs.harvard.edu/abs/1978Natur.271..334R} {271, 334}

\bibitem[\protect\citeauthoryear{{Rodr{\'\i}guez-Pascual}
  et~al.,}{{Rodr{\'\i}guez-Pascual} et~al.}{1997}]{Rodriguez:1997}
{Rodr{\'\i}guez-Pascual} P.~M.,  et~al., 1997, \mn@doi [\apjs]
  {10.1086/312996}, \href
  {https://ui.adsabs.harvard.edu/abs/1997ApJS..110....9R} {110, 9}

\bibitem[\protect\citeauthoryear{{Roming} et~al.,}{{Roming}
  et~al.}{2005}]{Roming:2005}
{Roming} P. W.~A.,  et~al., 2005, \mn@doi [\ssr] {10.1007/s11214-005-5095-4},
  \href {https://ui.adsabs.harvard.edu/abs/2005SSRv..120...95R} {120, 95}

\bibitem[\protect\citeauthoryear{{Salpeter}}{{Salpeter}}{1964}]{Salpeter:1964}
{Salpeter} E.~E.,  1964, \mn@doi [\apj] {10.1086/147973}, \href
  {https://ui.adsabs.harvard.edu/abs/1964ApJ...140..796S} {140, 796}

\bibitem[\protect\citeauthoryear{{Schlafly} \& {Finkbeiner}}{{Schlafly} \&
  {Finkbeiner}}{2011}]{Schlafly:2011}
{Schlafly} E.~F.,  {Finkbeiner} D.~P.,  2011, \mn@doi [\apj]
  {10.1088/0004-637X/737/2/103}, \href
  {https://ui.adsabs.harvard.edu/abs/2011ApJ...737..103S} {737, 103}

\bibitem[\protect\citeauthoryear{{Shakura} \& {Sunyaev}}{{Shakura} \&
  {Sunyaev}}{1973}]{Shakura:1973}
{Shakura} N.~I.,  {Sunyaev} R.~A.,  1973, \aap, \href
  {https://ui.adsabs.harvard.edu/abs/1973A&A....24..337S} {500, 33}

\bibitem[\protect\citeauthoryear{{Siegel}, {Porterfield}, {Balzer}  \&
  {Hagen}}{{Siegel} et~al.}{2015}]{Siegel:2015}
{Siegel} M.~H.,  {Porterfield} B.~L.,  {Balzer} B.~G.,   {Hagen} L. M.~Z.,
  2015, \mn@doi [\aj] {10.1088/0004-6256/150/4/129}, \href
  {https://ui.adsabs.harvard.edu/abs/2015AJ....150..129S} {150, 129}

\bibitem[\protect\citeauthoryear{{Starkey}, {Horne}  \& {Villforth}}{{Starkey}
  et~al.}{2016}]{Starkey:2016}
{Starkey} D.~A.,  {Horne} K.,   {Villforth} C.,  2016, \mn@doi [\mnras]
  {10.1093/mnras/stv2744}, \href
  {https://ui.adsabs.harvard.edu/abs/2016MNRAS.456.1960S} {456, 1960}

\bibitem[\protect\citeauthoryear{{Starkey} et~al.,}{{Starkey}
  et~al.}{2017}]{Starkey:2017}
{Starkey} D.,  et~al., 2017, \mn@doi [\apj] {10.3847/1538-4357/835/1/65}, \href
  {https://ui.adsabs.harvard.edu/abs/2017ApJ...835...65S} {835, 65}

\bibitem[\protect\citeauthoryear{{Stetson}}{{Stetson}}{1987}]{Stetson:1987}
{Stetson} P.~B.,  1987, \mn@doi [\pasp] {10.1086/131977}, \href
  {https://ui.adsabs.harvard.edu/abs/1987PASP...99..191S} {99, 191}

\bibitem[\protect\citeauthoryear{{Uttley}, {Edelson}, {McHardy}, {Peterson}  \&
  {Markowitz}}{{Uttley} et~al.}{2003}]{Uttley:2003}
{Uttley} P.,  {Edelson} R.,  {McHardy} I.~M.,  {Peterson} B.~M.,   {Markowitz}
  A.,  2003, \mn@doi [\apjl] {10.1086/373887}, \href
  {https://ui.adsabs.harvard.edu/abs/2003ApJ...584L..53U} {584, L53}

\bibitem[\protect\citeauthoryear{{Uttley}, {Wilkinson}, {Cassatella}, {Wilms},
  {Pottschmidt}, {Hanke}  \& {B{\"o}ck}}{{Uttley} et~al.}{2011}]{Uttley:2011}
{Uttley} P.,  {Wilkinson} T.,  {Cassatella} P.,  {Wilms} J.,  {Pottschmidt} K.,
   {Hanke} M.,   {B{\"o}ck} M.,  2011, \mn@doi [\mnras]
  {10.1111/j.1745-3933.2011.01056.x}, \href
  {https://ui.adsabs.harvard.edu/abs/2011MNRAS.414L..60U} {414, L60}

\bibitem[\protect\citeauthoryear{{Valenti} et~al.,}{{Valenti}
  et~al.}{2016}]{valenti:2016}
{Valenti} S.,  et~al., 2016, \mn@doi [\mnras] {10.1093/mnras/stw870}, \href
  {https://ui.adsabs.harvard.edu/abs/2016MNRAS.459.3939V} {459, 3939}

\bibitem[\protect\citeauthoryear{{Vasudevan} \& {Fabian}}{{Vasudevan} \&
  {Fabian}}{2009}]{Vasudevan:2009}
{Vasudevan} R.~V.,  {Fabian} A.~C.,  2009, \mn@doi [\mnras]
  {10.1111/j.1365-2966.2008.14108.x}, \href
  {https://ui.adsabs.harvard.edu/abs/2009MNRAS.392.1124V} {392, 1124}

\bibitem[\protect\citeauthoryear{{Vaughan}, {Edelson}, {Warwick}  \&
  {Uttley}}{{Vaughan} et~al.}{2003}]{Vaughan:2003}
{Vaughan} S.,  {Edelson} R.,  {Warwick} R.~S.,   {Uttley} P.,  2003, \mn@doi
  [\mnras] {10.1046/j.1365-2966.2003.07042.x}, \href
  {https://ui.adsabs.harvard.edu/abs/2003MNRAS.345.1271V} {345, 1271}

\bibitem[\protect\citeauthoryear{{Wheeler} \& {Chambers}}{{Wheeler} \&
  {Chambers}}{1992}]{Wheeler:1992}
{Wheeler} D.~J.,  {Chambers} D.~S.,  1992, {Understanding statistical process
  control}

\bibitem[\protect\citeauthoryear{{White} \& {Peterson}}{{White} \&
  {Peterson}}{1994}]{White:1994}
{White} R.~J.,  {Peterson} B.~M.,  1994, \mn@doi [\pasp] {10.1086/133456},
  \href {https://ui.adsabs.harvard.edu/abs/1994PASP..106..879W} {106, 879}

\bibitem[\protect\citeauthoryear{{Zel'dovich}}{{Zel'dovich}}{1964}]{Zeldovich:1964}
{Zel'dovich} Y.~B.,  1964, Soviet Physics Doklady, \href
  {https://ui.adsabs.harvard.edu/abs/1964SPhD....9..195Z} {9, 195}

\makeatother
\end{thebibliography}

\newpage
\appendix
\section{A tool for filtering ``dropouts'' from Swift UVOT data}\label{sec:app_map}

\subsection{Detector map from Galactic centre data}

As first noted by \cite{Edelson:2015}, \swift\ UVOT light curves exhibit occasional ``dropouts:'' fluxes that are anomalously low relative to the nearby data and apparently not intrinsic to the source.
Our earlier work indicated this is due to localised low sensitivity regions in the UVOT detector plane. We developed a methodology that used these dropouts in the AGN light curves to map and mask out these problematic regions \citep[e.g][for details]{Edelson:2015,Edelson:2017,Edelson:2019}.
However, because these observations only sparsely sample the detector plane, masks defined from them are optimised only for correcting the light curve of the target AGN.
Here we develop more detailed, generally applicable maps of the problematic detector regions and apply them to mask the \fair\  data, as discussed below.

The Galactic centre (Sgr A*) has been monitored with near-daily observations since early in the \swift\ mission \citep{Degenaar:2013}.
Each observation provides numerous sensitivity measurements scattered across the full UVOT field of view (FoV) due to the high stellar density of this field.
This provides a near-ideal dataset for this endeavour, as the FoV is sampled densely and relatively uniformly, as opposed to the AGN data, which sparsely measure only a portion of the detector plane.

Sgr~A* field star data were taken from 3202 observations from 2006-06-02 through 2018-04-22.  
Three hundred fifty-five field stars were tracked and their fluxes measured whenever they fell within the UVOT FoV, yielding a total of 531,132 measurements.
These data were screened and processed following the same procedures applied to the \fair\ data (Section~\ref{sec:uvot}). 
Additional screening eliminated any data points with large coincidence loss correction factors or flux uncertainties.
Light curves for each star in each filter are modelled by fitting a second-order polynomial to the neighbouring data points within a sliding window centred upon each measurement, iteratively rejecting outliers.
Final model values are recorded only for measurements that retain at least nine neighbours that constrain the light curve model without extrapolation.
To ensure model quality, light curves are rejected if they included too few data points ($<50$ after screening), do not fully represent variability in the data,
or if the median uncertainty of data used to define the light curve is large.
This eliminates all B and V light curves, as there are few Sgr A* observations in these filters.

To generate UVOT sensitivity maps, the fractional flux differences between measurements and light curve models are projected onto the detector plane.
The flux deviations are evaluated for a final sample of 136,718 data points with robust light curve model values and small measurement uncertainties. 
Preliminary detector coordinates for each measurement are obtained from the centroids of Gaussians fitted to the stars in raw images.
These coordinates are then corrected for offsets due to shift-and-add (S\&A) processing.
The median of the S\&A offsets applied to each image (which is typically 8--10\arcsec and can be determined from log files) is subtracted from the fitted coordinates to determine the detector position that most strongly influences each measurement.

\begin{figure*}
\includegraphics[width=1.\columnwidth]{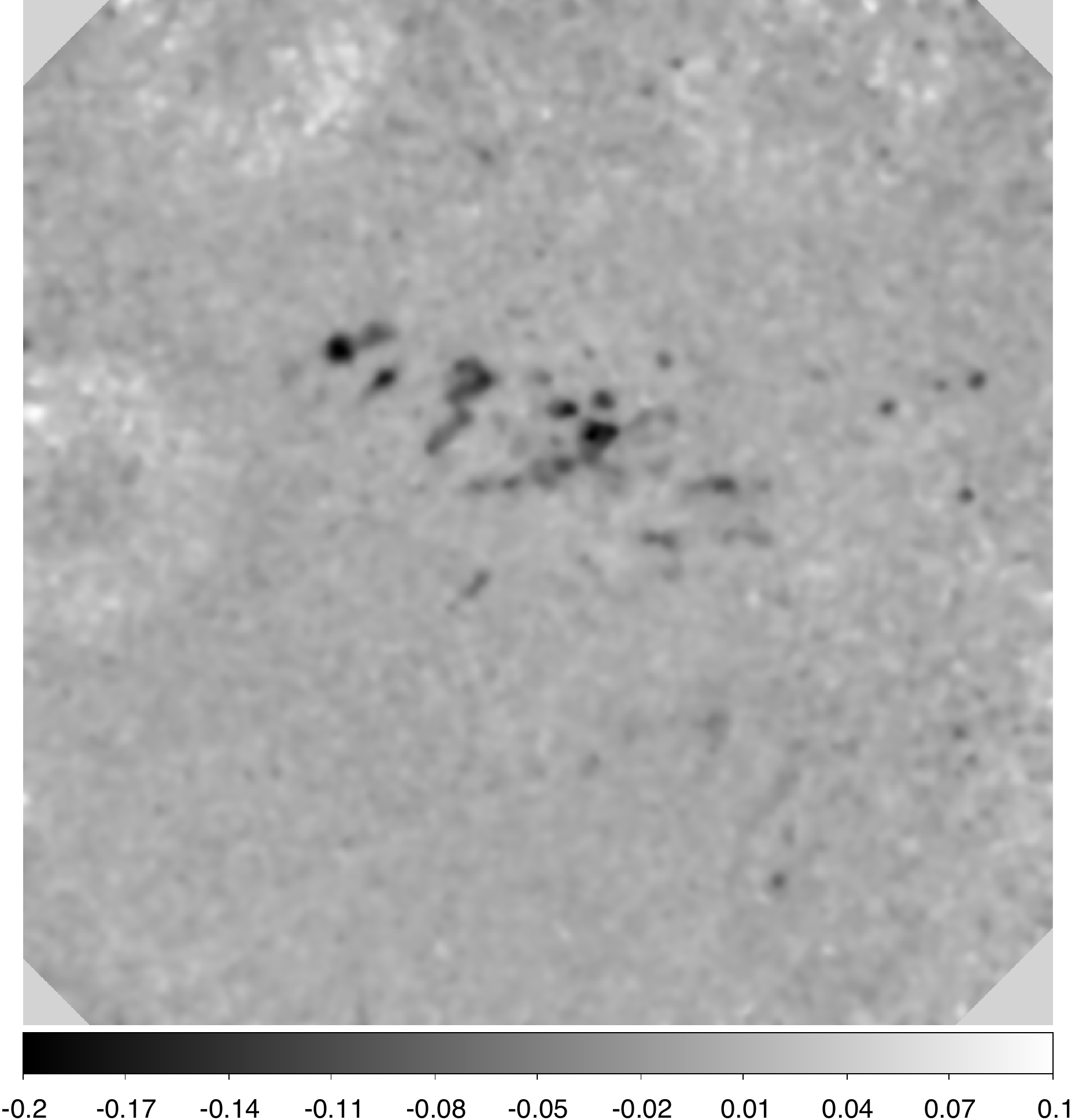}
\includegraphics[width=1.\columnwidth]{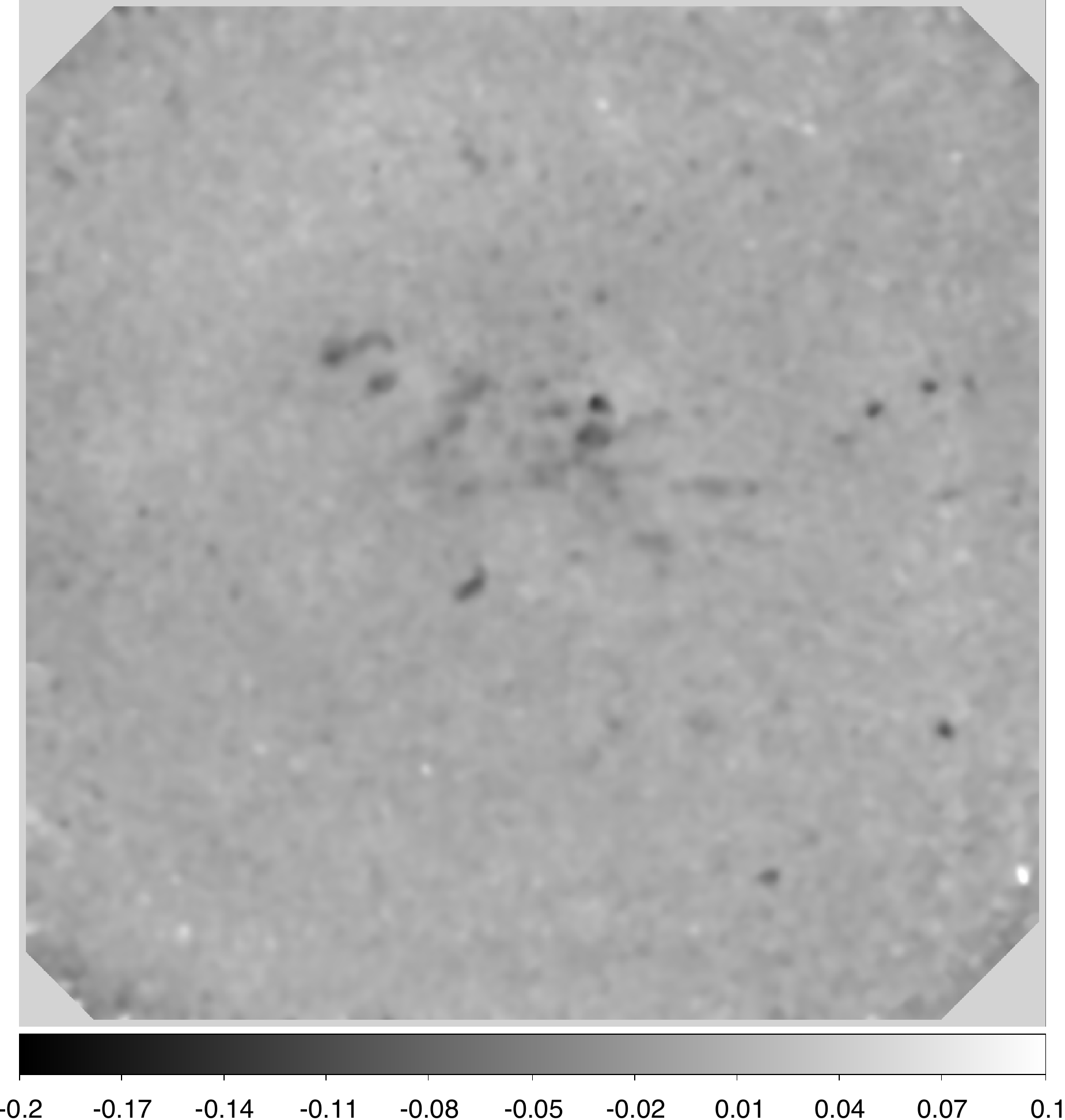}
\caption{Heat maps combining data from all UV filters (left) and U filter data (right), both smoothed with a 5\arcsec\ kernel.  
These are the final maps used to define the detector masks for the UV and optical bands.
The greyscale ranges of the panels are matched, showing that the effect of the low sensitivity regions is greater in the UV than in the optical.
The primary low sensitivity (dark) regions in the two maps line up, but the most extreme regions in one map are not the darkest regions in the other.
\label{fig:A1}}
\end{figure*}

The final sensitivity maps are produced by assigning to each 1\arcsec\ pixel (after 2$\times$ binning) the average of all nearby deviation measurements, weighted by distance from the pixel centre.
By applying Gaussian weighting functions with different kernel sizes, sets of maps are generated with varying amounts of smoothing.
Such sets are generated for each filter.
In addition, multi-filter maps that combine W2 and M2 or all three UV maps are generated to improve the sampling statistics, as there are strong similarities between these maps and fewer light curve deviation measurements are available in the UV filters.
Sample maps are shown in Figure~\ref{fig:A1}.

Field stars from the AGN IDRM campaigns are used to test the heat maps.
These provide a pool of independent measurements in six UVOT filters, with high-cadence monitoring and spatial sampling that spans the FoV.
The same procedure is followed to model the stellar light curves and select the highest quality data to use for measuring deviations from the models.
The test pool is limited to data within 4.5\arcmin\ of the FoV centre, as this is where 95\% of planned targets fall.
The final set of test data includes 1200--1600 measurements in each UV filter and 2900--5100 in the optical filters.
The flux deviations measured from these data correlate with the corresponding heat map values, thereby corroborating the mapping of detector sensitivity variations.

The strengths of these correlations are used to make the final heat map selection.
In each UV filter the deviations are found to correlate most strongly with the maps that combine Sag A* deviations in all three UV filters, whereas the optical measurements correlate better with the heat map defined from U-filter data.
As for the smoothing kernel choice, the correlation strength peaks with a kernel size of 5\arcsec\ for the UV filters and 4\arcsec\ in the optical bands.
However, the correlation coefficient differences between these kernels are mostly just a few thousandths, so a 5\arcsec\ kernel is selected for all maps to provide consistency across all filters.
The selected maps, shown in Figure~\ref{fig:A1}, will be made available for download in Zenodo.

Masks are defined to screen out data most strongly affected by regions of low sensitivity, as indicated by the heat maps.  
While the exact amplitude of the effect depends upon the filter and smoothing dilutes gradients and structure in the maps, thresholds may be used to identify where the effect is strongest.  
The thresholds are set by comparing the test pool deviation measurements to the corresponding heat map values at their detector locations.  
Masking thresholds are set at the highest heat map value for which 80\% of the test pool deviations are negative.  
The masks for W2, M2 and W1 are defined using threshold values of -0.01570, -0.01152, and -0.01100 (respectively) with the UV heat map, while the masks for U, B and V apply thresholds of -0.01333, -0.02641, and -0.02900 to the optical heat map.  
These masks can be downloaded as a single FITS file from Zenodo.  
To apply them, albeit without accounting for S\&A offsets in the data (as of June 2020), follow the UVOT small scale sensitivity check procedure\footnote{This procedure is described by the UVOT team on their \href{https://swift.gsfc.nasa.gov/analysis/uvot_digest/sss_check.html}{Small Scale Sensitivity page}.}, substituting the new masks as the LSSFILE parameter passed to the FTOOLS UVOT measurement routines (e.g., UVOTSOURCE).

\begin{figure*}
\includegraphics[width=2.\columnwidth]{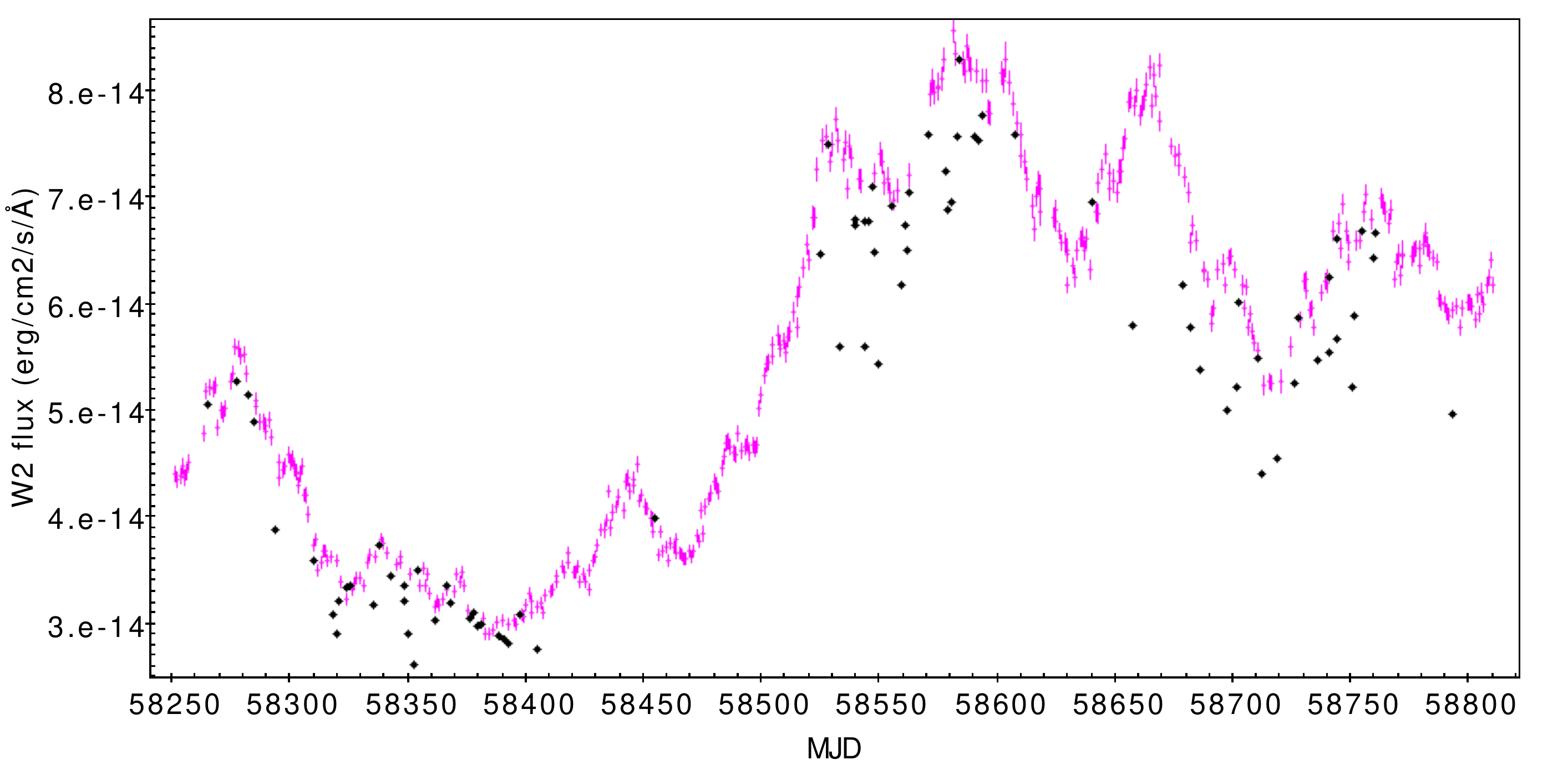}
\caption{\fair\ light curve in W2.  
The filled, coloured points with error bars are the final sample of W2 measurements while the smaller black points are ones that pass all other screening but are flagged by the mask of low sensitivity regions.
\label{fig:maskedF9lc}}
\end{figure*}

Figure~\ref{fig:maskedF9lc} shows the result when the W2 mask is applied to the \fair\ light curve: all clear dropouts and many other low data points are screened out, while very few screened points are consistent with neighbouring points in the light curve.
In total, 16.5\% of the measurements are masked (82 out of 497).
When the same mask is applied to the W2 test pool data, 13.7\% (222 out of 1622) data points are screened.  
The difference in percentage is due to the spatial distribution of measurements, as the test pool data are spread fairly evenly across the centre of the FoV whereas pointed observations at a single target over consecutive days tend to
fall in clusters on the detector.  
This spatial correlation is what causes the masked data points to be concentrated in specific time periods across the light curve.

\subsection{Detector map from M3 data}

As an additional independent check of our updated dropout filter mask, we reprocessed the photometry of the globular cluster M3 (NGC~5272) presented in \citet{Siegel:2015}.  We produced {\sc daophot} \citep{Stetson:1987} point-source photometry from the {\it raw} UVM2 spacecraft images, which are corrected for instrumental effects but not translated to a sky reference frame. This means that detections are in CCD pixel coordinates. After performing PSF photometry, we matched all the frames, calculated variability for each point source and searched the database, comprising 285,000 photometric measures of 5539 stars on 72 individual frames, for drops in flux greater than three times the photometric uncertainty.

This analysis is complicated by M3 being host to many RR Lyrae variable stars, which tend to dominate the UV detections.  Indeed, the data come from a survey designed to identify and characterise RR Lyrae stars, which have very large pulsations in the UV \citep[][and references therein]{Siegel:2015}.  This tends to produce both more photometric outliers, larger photometric outliers and a bias toward positive residuals (i.e., detections of RR~Lyrae when they are brighter than their mean flux).

Nonetheless we find that the M3 analysis yields results that are very similar to those obtained for the Galactic Centre
The M3 data reveal a distinct clustering of low flux measurements beginning slightly above the chip centre and extending toward the upper right portion of the chip.  
The amplitude of the dropouts tend to be significantly larger than seen in the heat maps (20--50\%) due to a combination of factors: the M3 deviation measurements combine intrinsic variability with detector bias, while the values in the heat map are diluted by measurements in the UVW1 filter, which have smaller deviation amplitudes and contribute the largest fraction of data in the map.  When considering just Sgr A$^*$ measurements in the UVM2 band, the largest flux deviation measured is 33.4\%.
A more detailed analysis will fit intrinsically variable light curves and measure deviation from them.  However, this preliminary analysis does confirm both methods and that this dropout effect is seen in point source photometry generated from all UVOT images, even photometry generated a {\it local} solution to the background sky level.

\section{Light curves of \fair}\label{sec:app_data}
Here we present the format for all the light curves used in this paper. The \swift/UVOT UV filters are noted as UVW2=W2, UVM2=M2 and UVW1=W1. The optical filters names are provided with two letters, where the first one corresponds to the filter name and the second one to the observatory. For example, {\sc BS} corresponds to $B$ band in \swift\ while {\sc BL} to $B$-band in \LCO. The full data set is available in a machine readable format.

The calibrated errors for the LCO light curves were obtained by adding in quadrature the additional scatter obtained in Section~\ref{sec:intercal}. These values for each filter are presented in Table~\ref{tab:uncertainties}. In general, the systematic uncertainty is a factor of $\sim3-5$ larger than the original error bars.

\begin{table}
	\centering
	\caption{Light curve file format for year-1 of the IDRM of \fair. The light curves show the inter-calibrated light curves, as described in Sec~\ref{sec:intercal}. This table is available in machine-readable format.}
	\label{tab:data_format}
	\begin{tabular}{cccc} 
	\hline
	  Filter & Time & Flux & Error\\
	   & MJD & mJy & mJy\\
	 \hline
    W2 & 58251.65954 & 5.454 & 0.089 \\
    W2 & 58252.45592 & 5.383 & 0.087 \\
    W2 & 58253.86214 & 5.430 & 0.084 \\
    W2 & 58254.58299 & 5.548 & 0.090 \\
    W2 & 58255.51821 & 5.405 & 0.093 \\
    $\cdots$&$\cdots$&$\cdots$&$\cdots$\\
    zL & 58522.04347 & 14.075 & 0.153\\
    zL & 58523.04287 & 14.143 & 0.153\\
    zL & 58523.43402 & 13.757 & 0.154\\
    zL & 58524.43295 & 14.113 & 0.154\\
    zL & 58529.03947 & 14.450 & 0.153\\
	 \hline
	 
	\end{tabular}
\end{table}

\begin{table}
	\centering
	\caption{Systematic error determination for the \LCO\ light curves. The mean error shows the values of the uncertainties from our data reduction pipeline. The Added Error shows the values obtained in Section~\ref{sec:intercal}. These were added to every data point in quadrature. The last column shows the ratio between the original average error bars and those after the inter-calibration process.}
	\label{tab:uncertainties}
\begin{tabular}{lccc}
\hline
Filter                           & Mean error & Added Error & Error ratio \\
                                 & mJy        & mJy         &             \\ 
\hline
$u^\prime$          & 0.032      & 0.183       & 5.76        \\ 
$B$          & 0.016      & 0.078       & 4.86        \\ 
$g^\prime$          & 0.013      & 0.056       & 4.34        \\ 
$V$          & 0.019      & 0.073       & 3.81        \\ 
$r^\prime$          & 0.020      & 0.065       & 3.27        \\ 
$i^\prime$          & 0.025      & 0.095       & 3.83        \\
$z_s$          & 0.030      & 0.152       & 5.03       \\
\hline
\end{tabular}
\end{table}

%
\section{{\sc cream} fit light curves to a face-on accretion disc}\label{sec:app_cream}
\begin{figure*}
\begin{center}
	\includegraphics[trim=3cm 6.5cm 1.5cm 0cm, clip, width=2.1\columnwidth]{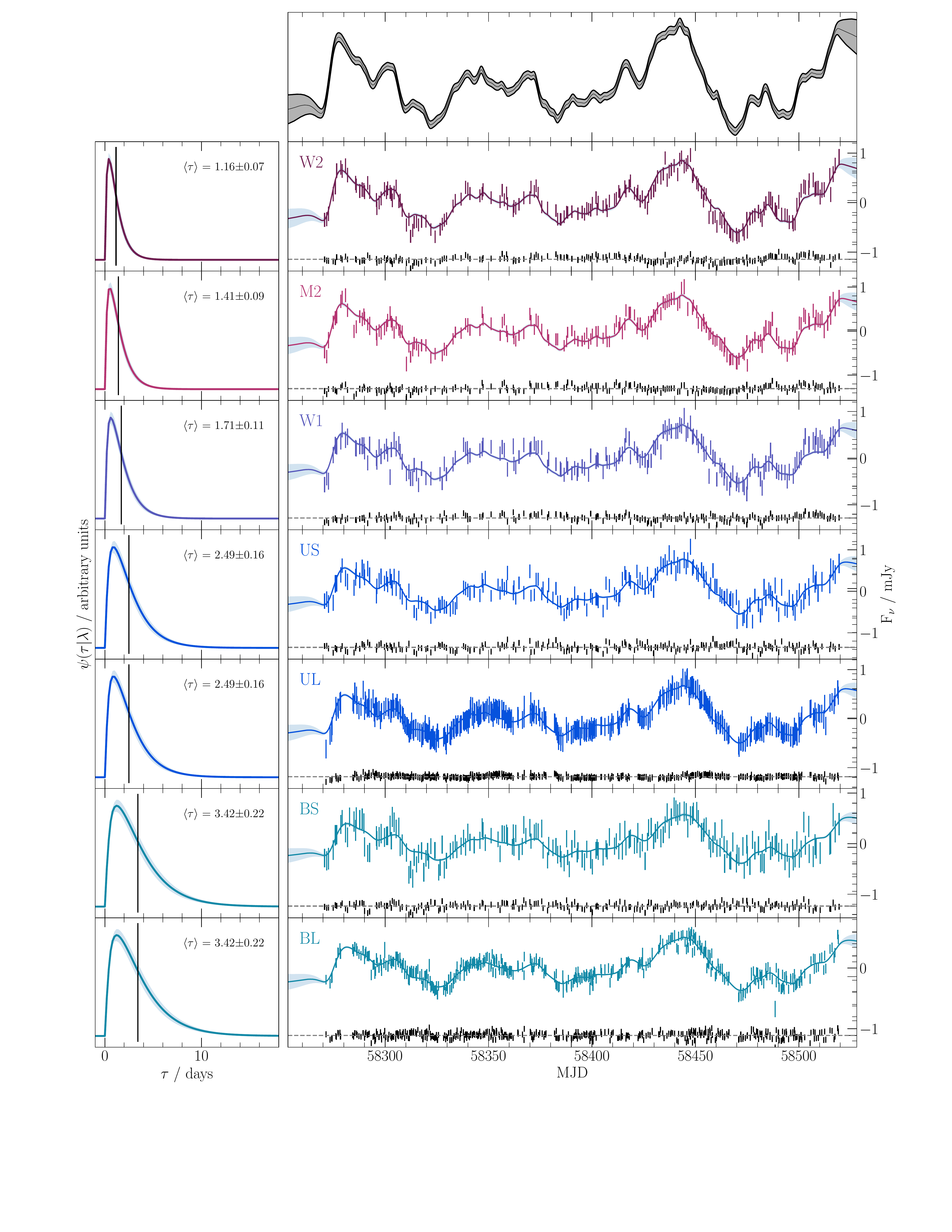}
    \caption{Reverberation model fit with {\sc cream} to a face-on accretion disc and a $T\propto R^{-3/4}$ temperature profile for \fair. Top panel shows the inferred driving light curve. Left panels show the delay distribution for each band. Right panel shows the photometry after the quadratic detrend and the best fit. The black points around the dotted grey line show the residuals. All grey envelopes represent the 1$\sigma$ confidence interval.}
    \label{fig:cream_fit1}
\end{center}
\end{figure*}

\begin{figure*}
    \begin{center}
	\includegraphics[trim=3cm 6.5cm 1.5cm 0cm, clip, width=2.1\columnwidth]{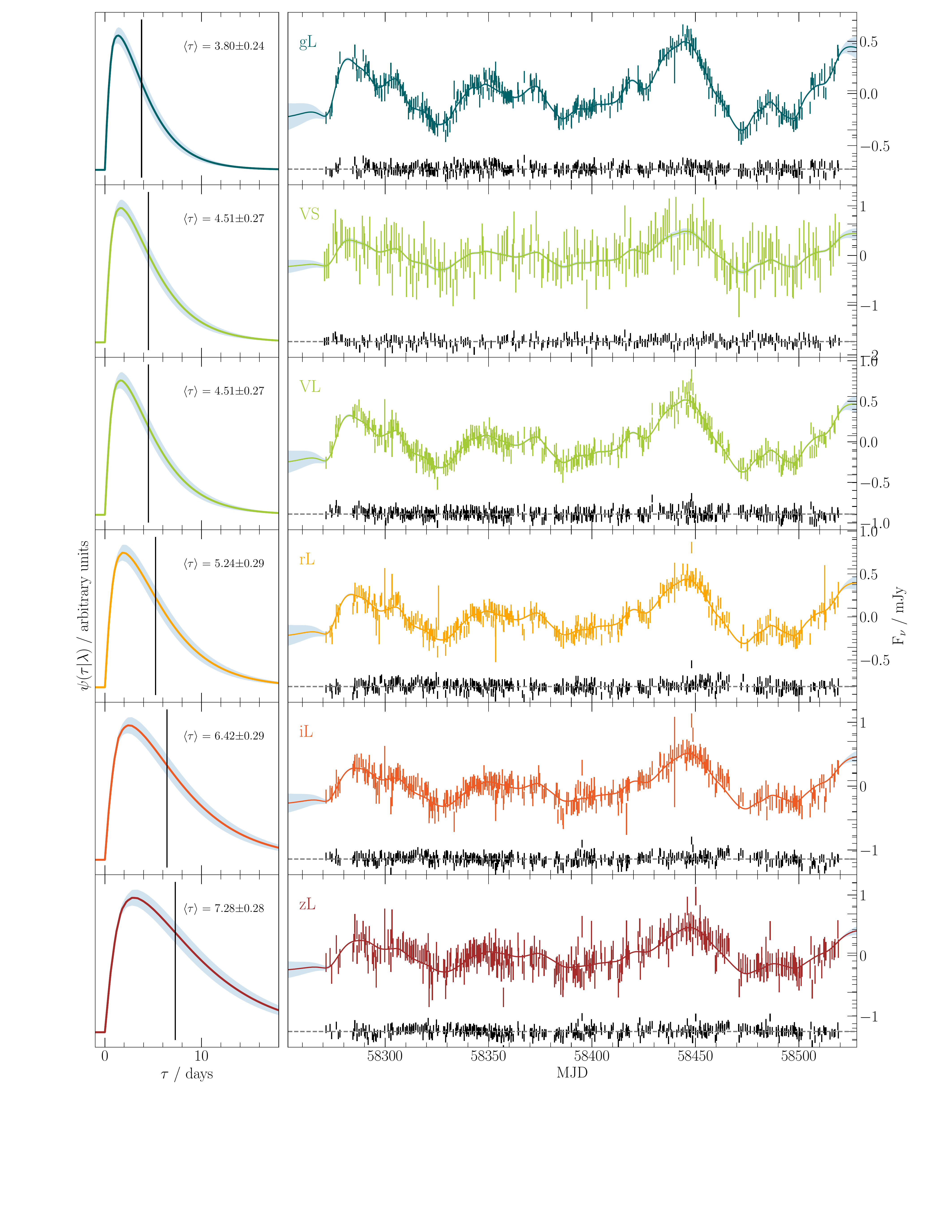}
    \caption{Reverberation model fit with {\sc cream} to a face-on accretion disc and a $T\propto R^{-3/4}$ temperature profile for \fair. Left panels show the delay distribution for each band. Right panel shows the photometry after the quadratic detrend and the best fit. The black points around the dotted grey line show the residuals. All grey envelopes represent the 1$\sigma$ confidence interval.}
    \label{fig:cream_fit2}
    \end{center}
\end{figure*}

Figures \ref{fig:cream_fit1} and \ref{fig:cream_fit2} show the best fit from {\sc cream} to every \swift\ and \LCO\ band. 

\section{Parameter distribution for the lag spectrum fit}
\begin{figure*}
    \begin{center}
	\includegraphics[trim=0cm 0.5cm 0.5cm 0cm, clip, width=2.1\columnwidth]{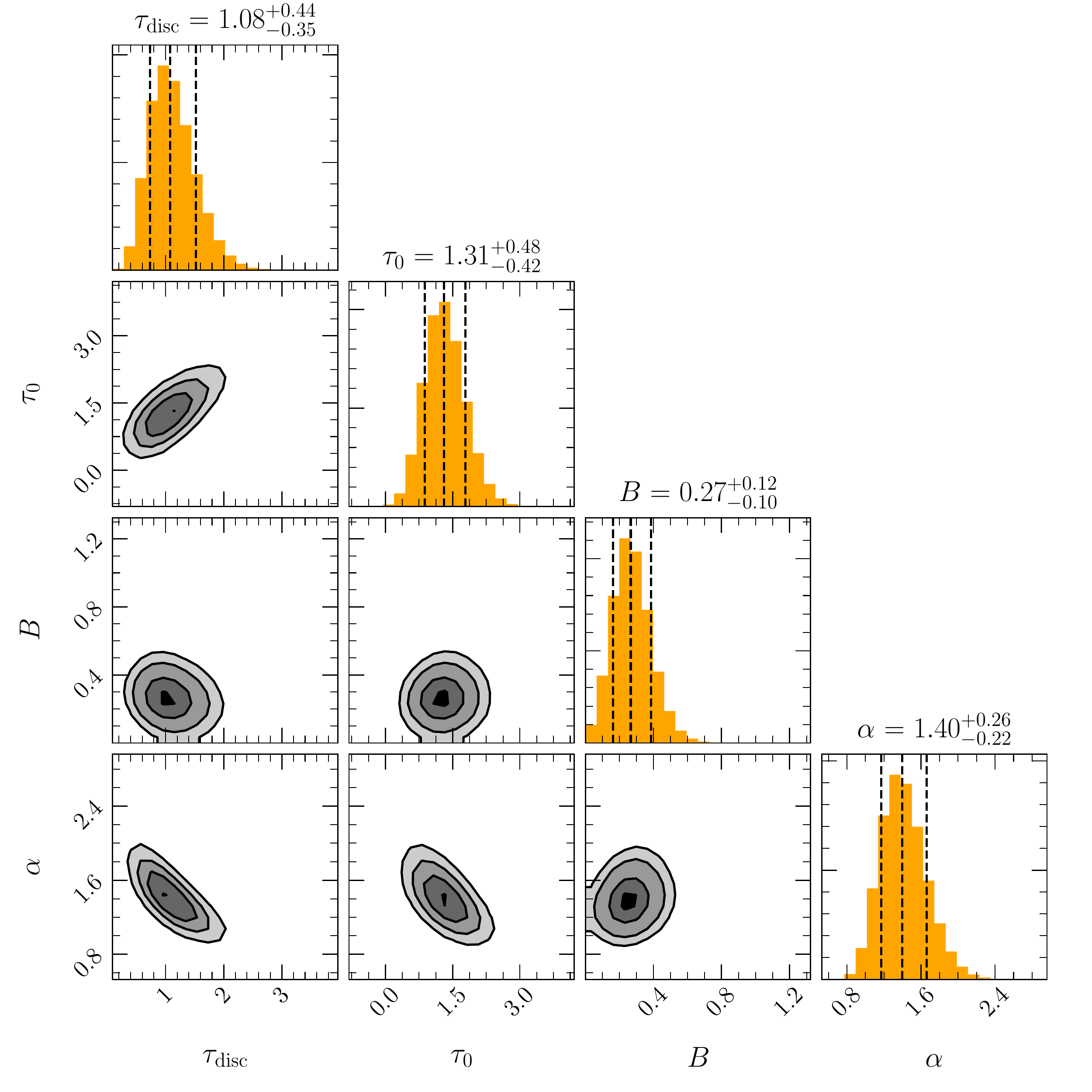}
    \caption{Posterior probability distributions for the accretion disc parameters. Colour scale contours show the joint probability for every combination of parameters. Contours represent the 0.5$\sigma$, 1$\sigma$ , 2$\sigma$ and 3$\sigma$ levels. Marginal posterior distributions are shown as histograms with the median and 1$\sigma$ marked as dashed lines.}
    \label{fig:A2}
    \end{center}
\end{figure*}

We present the joint and marginal posterior distributions of the lag spectrum fit (Fig.~\ref{fig:A2}) described in Sec.~\ref{sec:dce}. We used {\sc corner.py} \citep{corner} to visualise the MCMC chains.



\bsp	
\label{lastpage}
\end{document}